\documentstyle[emulateapj,amssym]{article}

\newcommand{\logg}{$\log\,g$}
\newcommand{\teff}{$T_{\rm eff}$}
\newcommand{\vsini}{$v \sin i$}
\newcommand{\stromu}{Str\"{o}mgren $u$}

\begin{document}

\received{}
\revised{}
\accepted{}

\lefthead{Grundahl, Catelan, Landsman, Stetson, \& Andersen}
\righthead{The ``Jump" in Str\"{o}mgren $u$, Low Gravities, and Levitation}

\submitted{{\rm The Astrophysical Journal}, accepted: May 7, 1999}

\singlespace

\title{Hot Horizontal-Branch Stars: The Ubiquitous Nature of 
the ``Jump" in Str\"{o}mgren \lowercase{$u$},\\Low Gravities, and the Role 
of Radiative Levitation of Metals\,\,\altaffilmark{1,}\altaffilmark{2}} 

\author{F.~Grundahl \altaffilmark{3,}\altaffilmark{4,}\altaffilmark{5}}
\authoraddr{Frank.Grundahl@hia.nrc.ca}
\author{M.~Catelan \altaffilmark{6,}\altaffilmark{7,}\altaffilmark{8}}
\authoraddr{catelan@gsfc.nasa.gov}
\author{W.~B.~Landsman \altaffilmark{9} }
\authoraddr{landsman@mpb.gsfc.nasa.gov}
\author{P.~B.~Stetson \altaffilmark{3,}\altaffilmark{5} }
\authoraddr{Peter.Stetson@hia.nrc.ca}
\and
\author{M.~I.~Andersen \altaffilmark{10}}
\authoraddr{andersen@not.iac.es}

\altaffiltext{1}{Based on observations made with the Nordic Optical 
   Telescope, operated on the island of La Palma jointly by Denmark, 
   Finland, Iceland, Norway, and Sweden, in the Spanish Observatorio 
   del Roque de los Muchachos of the Instituto de Astrofisica de 
   Canarias.}
\altaffiltext{2}{Based on observations obtained with the Danish 1.5-m 
   telescope at the European Southern Observatory, La Silla, Chile.}
\altaffiltext{3}{
   Dominion Astrophysical Observatory, Herzberg Institute of Astrophysics,
   National Research Council, 5071 W. Saanich Road, Victoria, BC V8X 4M6, 
   Canada; Frank.Grundahl@hia.nrc.ca, Peter.Stetson@hia.nrc.ca
   }
\altaffiltext{4}{
   University of Victoria, Department of Physics \& Astronomy,
   PO Box 3055, Victoria, BC V8W 3P7, Canada
   }
\altaffiltext{5}{
   Guest user, Canadian Astronomy Data Centre, which is
   operated by the Dominion Astrophysical Observatory for the Canadian
   National Research Council's Herzberg Institute of Astrophysics}
\altaffiltext{6}{
   NASA Goddard Space Flight Center, Code 681, Greenbelt,
   MD 20771, USA; catelan@stars.gsfc.nasa.gov
   }
\altaffiltext{7}{
   Hubble Fellow
   }
\altaffiltext{8}{
   Visiting Scientist, Universities Space Research Association
   }
\altaffiltext{9}{
   Raytheon ITSS, NASA Goddard Space Flight Center, Code 681, Greenbelt,
   MD 20771, USA; landsman@mpb.gsfc.nasa.gov
   }
\altaffiltext{10}{
   Nordic Optical Telescope, Apartado 474, E-38700 Santa Cruz 
   de La Palma, La Palma, Spain; andersen@not.iac.es
   }

\begin{abstract}
A ``jump" in the blue horizontal-branch (HB) distribution in the $V$, $u-y$
color-magnitude diagram has recently been detected in the globular cluster
(GC) M13 (NGC~6205) by Grundahl, VandenBerg, \& Andersen (1998). Such an
effect is morphologically best characterized as a discontinuity in the $u$,
$u-y$ locus, with stars in the range $11,\!500~{\rm K} \lesssim T_{\rm eff}
\lesssim 20,\!000$~K deviating systematically from (in the sense of
appearing brighter and/or hotter than) canonical zero-age HB models.  

In this article, we present Str\"omgren $u$, $y$ photometry of
fourteen globular clusters obtained with three different telescopes  (ESO
Danish, Nordic Optical Telescope, and the {\em Hubble Space Telescope}),
and demonstrate that the jump in Str\"omgren $u$ is present in 
every GC whose HB extends beyond $T_{\rm eff} \gtrsim 11,\!500$~K, 
irrespective of metallicity, mixing history on the red giant branch (RGB),
or any known parameter characterizing our sample of GCs. We thus suggest
that the $u$-jump is a ubiquitous feature, intrinsic to {\em all} HB stars
hotter than $T_{\rm eff} \simeq 11,\!500$~K.

We draw a parallel between the ubiquitous nature of the $u$-jump and 
the well-known problem of low measured gravities among blue-HB stars 
in globular clusters and in the field. We note that the ``gravity 
jump" occurs over the same temperature range as the $u$-jump, and 
also that it occurs in every metal-poor cluster for which gravities 
have been determined---again irrespective of metallicity, 
mixing history on the RGB, or any known parameter characterizing 
the surveyed GCs. Furthermore, we demonstrate that the $u$-jump 
and the gravity-jump are connected on a {\em star-by-star basis}. 
We thus suggest that the two most likely are different manifestations 
of one and the same physical phenomenon. 

We present an interpretative framework which may be capable of 
simultaneously accounting for both the $u$-jump and the gravity-jump. 
Reviewing spectroscopic data for several field blue-HB stars, as 
well as two blue-HB stars in NGC~6752, we find  evidence that 
radiative levitation of elements heavier than carbon and nitrogen 
takes place at $T_{\rm eff} \gtrsim 11,\!500$~K, {\em dramatically} 
enhancing the abundances of such heavy elements in the atmospheres 
of blue-HB stars in the ``critical" temperature region. 
We argue that model atmospheres which take diffusion effects 
into account are badly needed, and will likely lead to better 
overall agreement between canonical evolutionary theory and the 
observations for these stars. 

\end{abstract}


\newpage
\section{Introduction}

Galactic globular clusters (GCs) are the oldest known objects for which 
accurate ages can be derived. For this reason, they play a major role 
in posing a lower limit to the age of the Universe, thus constraining 
cosmological models (e.g., van den Bergh 1992; Bolte \& Hogan 1995; 
VandenBerg, Bolte, \& Stetson 1996; Chaboyer et al.\ 1996; Mould 1998) 
and scenarios for the early formation history of the Galaxy and its 
nearby companions (e.g., Eggen, Lynden-Bell, \& Sandage 1962; Mironov \& 
Samus 1974; Searle \& Zinn 1978; Zinn 1980, 1993; Brocato et al.\ 1996; 
Buonanno et al.\ 1998). 

From an observational point of view, for reliable GC ages to be determined 
it is extremely important that the Population~II distance scale be 
accurately established (Renzini 1981)---a task which has thus far not 
been successfully accomplished, even with the advent of {\sc Hipparcos} 
(e.g., Catelan 1998; Koen \& Laney 1998; Carretta et al.\ 1999a). From 
a theoretical point of view, 
it is crucial that the color-magnitude diagrams (CMDs) of GCs be accurately 
reproduced by theoretical isochrones and synthetic CMDs, so that the stellar 
structure and evolution models, as well as the model atmospheres used to 
transfer the predicted $\log\,L$ and $T_{\rm eff}$ values into observed 
magnitudes and colors, can be relied upon for ages to be determined from 
the observations (e.g., VandenBerg et al.\ 1996;  Salaris, Degl'Innocenti, 
\& Weiss 1997; VandenBerg \& Irwin 1997; Cassisi et al.\ 1999; VandenBerg 
1999). 

Of primary interest for these purposes are the main-sequence (MS) and 
horizontal-branch (HB) evolutionary phases. More specifically, both the 
MS turnoff luminosity and the HB morphology are sensitive to 
age, with the former being the standard clock for GC age determination 
(Iben \& Renzini 1984). The horizontal part of the HB in the $V$, $\bv$ 
plane, covering the RR Lyrae instability strip, most of the red HB and 
part of the blue HB [$(\bv)_0 \gtrsim 0.1$~mag], is the primary 
Population~II ``standard candle" (e.g., Gratton 1998). Though still 
quite uncertain as an age-derivation method, HB morphology in external 
galaxies is increasingly being used to place constraints on their ages and 
star formation histories (e.g., Da~Costa et al.\ 1996; Geisler et al.\ 1998). 
This highlights the importance of adequately interpreting the physical 
properties of HB stars in Galactic GCs and in the field. 

This notwithstanding, there are several long-standing open problems 
in the interpretation of observed HBs which have yet to find widely  
accepted explanations. Among these, we may quote:

 1.~The ``Oosterhoff-Arp-Sandage period-shift effect," which affects 
       the pulsation properties of RR Lyrae variables (Oosterhoff 1939; 
       Arp 1955; Sandage 1981). For recent discussions, see, for   
       instance: Smith (1995); Caputo (1998); Catelan, Sweigart, \& 
       Borissova (1998); Clement \& Shelton (1999); De Santis \& Cassisi 
       (1999); Layden et al.\ (1999); and Sweigart (1997a, 1997b); 

 2.~The ``second-parameter phenomenon": besides metallicity [Fe/H] 
       (the ``first parameter"; Sandage \& Wallerstein 1960), there 
       must be {\em at least} one additional parameter determining  
       HB morphology (Sandage \& Wildey 1967; van den Bergh 1967). 
       For recent discussions, see, e.g., Chaboyer, Demarque, 
       \& Sarajedini (1996); Stetson, VandenBerg, \& Bolte (1996);  
       Buonanno et al.\ (1997, 1998); Kraft et al.\ (1998); Sweigart 
       (1997a, 1997b); and Sweigart \& Catelan (1998); 

 3.~HB ``bimodality" (Harris 1974) and ``gaps" (Newell 1973; 
       Lee \& Cannon 1980). For recent discussions, see 
       Catelan et al.\ (1998); Ferraro et al.\ (1998); Caloi (1999); 
       and Piotto et al.\ (1999);  

 4.~The origin and nature of blue subdwarf (sdB) stars in the 
       field (Greenstein 1971) and in GCs (Caloi et al.\ 1986; Heber 
       et al.\ 1986). Blue subdwarfs are often called ``extreme" (or 
       ``extended") HB (EHB) stars (Greenstein \& Sargent 1974). 
       Following original suggestions by Caloi (1989) and Greggio \& 
       Renzini (1990), these stars and their progeny are now widely 
       believed (e.g., J{\o}rgensen \& Thejll 1993; Bressan, Chiosi, \& 
       Fagotto 1994; Dorman, O'Connell, \& Rood 1995;  Yi, Demarque, 
       \& Oemler 1998) to be the main contributors to the ultraviolet 
       light emanating from elliptical galaxies and the bulges of 
       spirals that is commonly referred to as the ``UV-upturn 
       phenomenon" (Code 1969). For recent discussions on the origin 
       and evolution of EHB stars, see D'Cruz et al.\ (1996); Rood, 
       Whitney, \& D'Cruz (1997); and Sweigart (1997b);  

 5.~Unexpectedly low surface gravities, as inferred from 
       fitting Balmer-line profiles for both field (Saffer et al.\ 
       1994, 1997; Mitchell et al.\ 1998) and GC (Crocker, Rood, \& 
       O'Connell 1988; de Boer, Schmidt, \& Heber 1995; Moehler, 
       Heber, \& de Boer 1995; Moehler, Heber, \& Rupprecht 1997; 
       Bragaglia et al.\ 1997) blue-HB (BHB) stars; 

 6.~The anomalous ``jump" in the $V$, $u-y$ CMD at the BHB 
       region, recently detected by Grundahl, VandenBerg, \& 
       Andersen (1998) in their study of the Galactic GC 
       M13 (NGC~6205).

The above problems are probably somewhat intertwined, and remain 
essentially open. It is thus clear that much needs to be 
accomplished for a comprehensive understanding of the physical 
properties of HB stars to be achieved. Unless this 
is properly done, it will remain dubious whether such stars can 
be reliably employed to determine distances and ages, and hence 
to constrain Cosmology and models for the formation history of 
galaxies. 

Our goal, in the present article, is to address the last two issues 
listed above: the low $\log\,g$ values measured for BHB stars, 
and the Grundahl et al.\ (1998) ``jump." We shall demonstrate that:

 1.~The $u$-jump is a ubiquitous feature, likely present in every 
       single metal-poor GC which hosts HB stars with 
       $T_{\rm eff} \gtrsim 11,\!500$~K;

 2.~The $u$-jump and the similar feature present in $\log\,g$, 
       $\log\,T_{\rm eff}$ diagrams are probably different 
       manifestations of the same physical phenomenon, and 
       intrinsic to all BHB stars, whether in the field or in GCs;

 3.~The physical reason for the occurence of the jumps in $u$ 
       and $\log\,g$ is most likely radiative levitation of 
       elements heavier than carbon and nitrogen into the stellar 
       atmosphere, rather than a stellar interior/evolution effect.

In the next section, we describe the observations, using three 
different telescopes, which have led to the compilation of our 
large database of CMDs in the Str\"omgren system. Our adopted 
data reduction procedures are also described in \S2. In \S3, 
we demonstrate that the jump in $u$ is a ubiquitous feature, 
occurring in all studied GCs at essentially the same location in 
$T_{\rm eff}$, irrespective of any parameters characterizing 
the globulars; in \S4, we demonstrate that there is a strong 
correlation between the jump in $u$ and the low 
gravities found among BHB stars in GCs and the field; in 
\S5, we address what constraints these conclusions pose on 
non-canonical models which have been proposed to account for the 
gravity anomalies; in \S6, we point out that radiative levitation 
of elements heavier than carbon and nitrogen is well documented 
among both field and GC BHB stars lying in the ``critical" 
$T_{\rm eff}$ range where the jump takes place, and argue 
that model atmospheres with dramatically enhanced abundances of 
such heavy elements (as observed) may be able to explain the 
failure of canonical models to reproduce both the bright $u$ 
magnitudes and the low measured gravities. Finally, in \S7 
we provide a summary of our results. Some consequences of our 
proposed scenario are also laid out, as are our concluding remarks.

\section{Observations and Data Reduction}

The observations reported in this work have been 
collected from the Nordic Optical Telescope (NOT), 
the Danish~1.54m telescope on La Silla and the {\em Hubble 
Space Telescope} archive (M13).
The ground-based data were obtained 
in the Str{\"o}mgren $u$ and $y$ filters, whereas the HST   
observations made use of their close WFPC2 analogs: the 
F336W and F555W filters, respectively. 

Data from NOT were collected during observing runs in 1995, 
1997 and 1998. Stars from the lists of Olsen (1983, 1984) and 
Schuster \& Nissen (1988) were observed on two nights in 1995 
and four nights in 1998 under photometric conditions, to 
derive the transformation between the instrumental magnitudes 
and the standard system. The data for M13 have been described 
in Grundahl et al.\ (1998). For M3 (NGC~5272), M5 (NGC~5904), 
M12 (NGC~6218), M15 (NGC~7078), M56 (NGC~6779), and 
M92 (NGC~6341), the data were obtained using a thinned 
AR coated $2048 \times 2048$ 
pixel CCD camera, with 0\farcs11 pixelsize, thus covering 
approximately 3.75 arcminutes on a side. Most of the observations 
were obtained using tip/tilt correction (the HiRAC camera) and 
the FWHM of nearly all our images ranged between 0\farcs45 and 
1\farcs0. There was no significant variation of the point spread
function (PSF) over the field of view. In M3 and M92 we observed 
two overlapping fields, with one field centered on the cluster 
center to ensure a large sample of HB and red-giant branch (RGB) 
stars. For M12 and M56 our fields were centered on the cluster 
center. The data for M5, M12, M15 and M56 were obtained under 
non-photometric conditions and were consequently not calibrated.

The data from the Danish~1.54m telescope were collected during 
two observing runs in May and October of 1997. For both runs we used 
the Danish Faint Object Spectrograph and Camera (DFOSC) equipped with 
a thinned, AR coated  2048 $\times$ 2048 pixel CCD camera. The 
field covered was approximately 11 arcminutes in diameter.  During the 
October observing run data were collected for NGC~288, NGC~1851, 
M2 (NGC~7089), M79 (NGC~1904) and NGC 6752, all of which were 
observed on several photometric nights; approximately 
150 different standard stars again from the lists of Olsen (1983, 
1984) and Schuster \& Nissen (1988) were also observed. The data for 
NGC 6397 and M30 (NGC~7099) were collected during 
the observing run in May 1997, and only a small fraction of these 
are used for this paper. The data for these two clusters have not 
yet been calibrated. For all the observations the seeing ranged 
between 1\farcs3 and 2\farcs2.  As most of the stars used as standard 
stars were rather bright ($V = 8-10$~mag), the telescopes (NOT and 
the Danish 1.54m) were defocused during these observations in order 
to avoid saturating the CCD.

All photometric reductions of the cluster frames were done using 
the suite of programs developed by PBS: {\sc daophot}, 
{\sc allstar}, {\sc allframe} and {\sc daogrow} (see Stetson 1987, 
1990, 1994). Flat 
fields were obtained on each night during evening and morning 
twilight. Photometry for the defocused standard stars was derived 
using large-aperture photometry. Based on the frame--to--frame
scatter for the bright stars in the clusters with calibrated 
photometry we estimate that the errors in the photometric zeropoints
are below 0.02~mag for the observations from NOT, and less than 
0.03~mag for the data from ESO. The larger errors for the ESO data
are due to the poorer seeing encountered during the observations, 
which makes the estimation of the aperture corrections in crowded
fields more uncertain. Of the clusters studied, M2, M3, M13, M79, 
M92, NGC~288, NGC~1851 and NGC~6752 have data from photometric nights. 

The HST data for M13 were retrieved from the Canadian 
Astronomical Data Center (CADC) in Victoria, and reduced with 
{\sc daophot}, {\sc allstar} and {\sc allframe}. We have not 
calibrated these data since our purpose with their inclusion in 
this work was to check whether or not the $u$-band jump was present.
The reductions followed the standard reduction procedures used by
team-members for stellar photometry with HST (see, e.g., Stetson et 
al.\ 1998, 1999).

\section{The Ubiquitous Nature of the ``Jump" in 
Str\"omgren \lowercase{$u$}}

\subsection{The $u$-Jump as a Ubiquitous Feature}

The jump in Str\"omgren $u$ was first detected very 
recently by Grundahl et al.\ (1998) in their photometric study of 
M13. As an explanation for the effect, Grundahl et al.\ tentatively 
suggested that ``helium mixing" models (see \S5 below) might 
account for their observations. 

During the reduction of Str{\"o}mgren data from other observing 
runs, it was found that the $u$--jump was present in all clusters
with a sufficiently blue HB. Given the potentially dramatic 
implications that mixing of helium 
into the envelopes of HB stars would have upon the Pop.~II distance 
scale and GC ages, we decided to undertake a comprehensive 
and systematic study of CMDs 
for all our observed GCs. Here we restrict our discussion to the $u$ 
and $y$ bandpasses, since we have found that the jump detected by 
Grundahl et al.\ (1998) is definitely most pronounced when the $u$, 
$u-y$ plane is employed.\footnote{We suspect from our observations 
that the effects of the $u$-jump may also be present in the other 
Str{\"o}mgren filters, although to a much smaller extent.}

%
\begin{figure*}[b]
 \figurenum{1}
 \plotfiddle{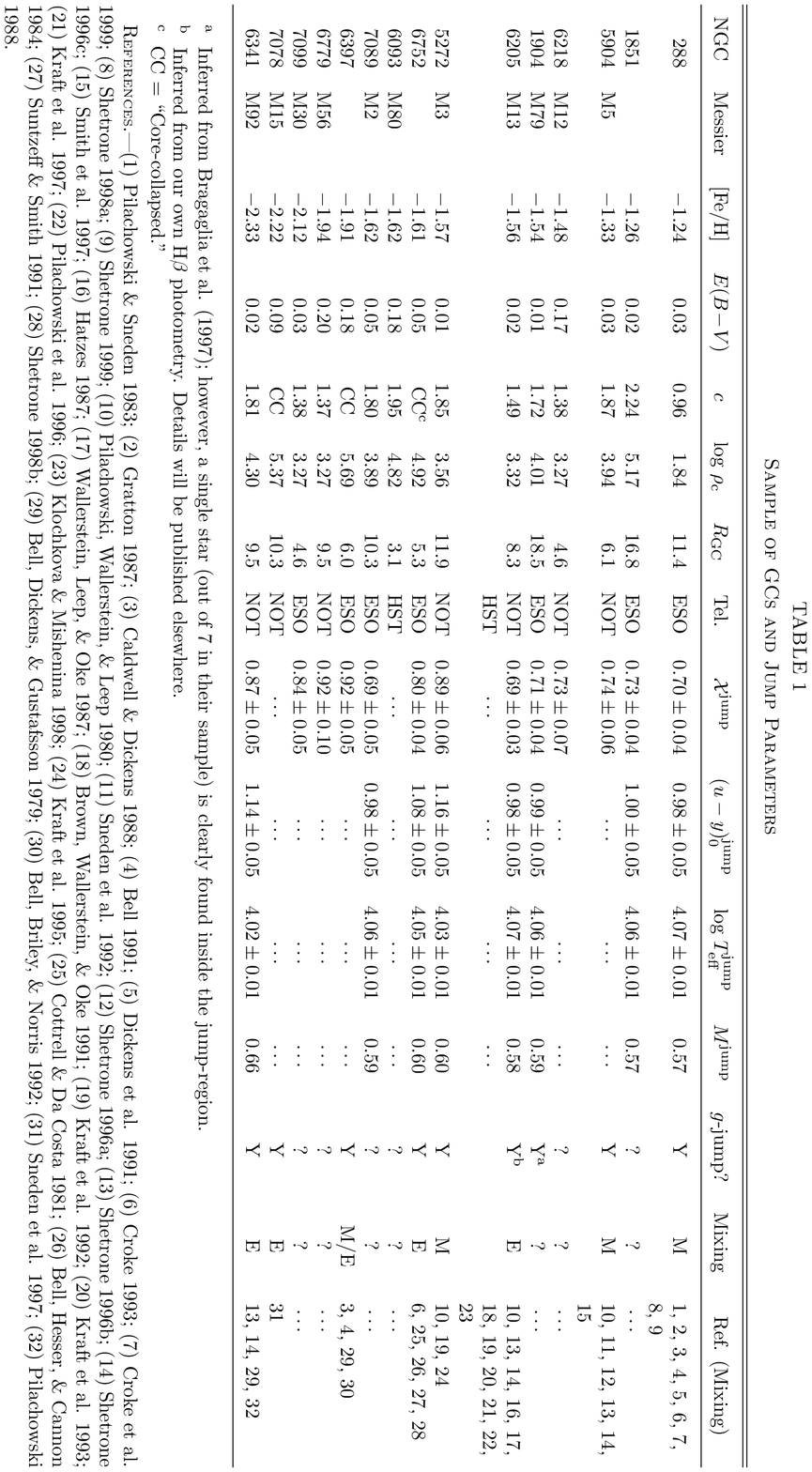}{3.85in}{90}{70}{70}{266}{0}
\end{figure*}

Table~1 shows the GC data set employed in this paper. In column~1, 
the cluster NGC number is given, as provided in ``The New General 
Catalogue of Nebulae and Clusters of Stars"; in 
column~2, the cluster name in Messier's catalogue is shown. 
The cluster metallicity [Fe/H], reddening $E(\bv)$, core concentration 
$c$, log of the central density 
$\rho_{\rm c}$ (in $L_{\sun}\,{\rm pc}^{-3}$), 
Galactocentric distance $R_{\rm GC}$ (in kpc) are given in 
columns~3 through 7 (data from Harris 1996). The telescope 
employed for the observations is given in column~8. 
In columns~9, 10, 11 and 12, $\cal{X}^{\rm jump}$, 
$(u-y)_0^{\rm jump}$, 
$T_{\rm eff}^{\rm jump}$ and mass location $M^{\rm jump}$ (in 
mag units, degrees Kelvin and $M_{\sun}$, respectively) of the 
low-temperature ``cutoff" of the jump in Str\"omgren $u$, 
estimated as described in \S3.2, are given. 
In column~13, information about the 
presence of a corresponding ``gravity jump" is supplied 
(``Y"~=~``Yes"; ``N"~=~``No"; question marks indicate clusters 
whose blue HBs have not been surveyed for $\log\,g$ values 
as of this writing). Finally, an estimate of the degree of 
mixing among cluster RGB stars (``E"~=~``Extensive"; 
``M"~=~``Moderate") is provided in column~14, based 
on the spectroscopic data from the references quoted in the final 
column. Again, question marks indicate clusters for which data 
are either not available or insufficient to reach any conclusion. 
In several cases, it is clear that (further) spectroscopic 
studies would be very helpful.  Note also that we have added 
M80 (NGC~6093) to our sample, since the photometry by Ferraro
et al.\ (1998) (their Fig. 5) clearly illustrates that in the 
F336W and F555W filters this cluster has an HB morphology very
similar to M13's. Thus we claim that this cluster shows the 
jump as well. 

Figure~1 shows a mosaic plot with the $u$, $u-y$ CMDs for the 
fourteen GCs in the present sample (M13 is shown twice, the data
from the HST being plotted separately from the NOT data). 
The CMDs are plotted in order of decreasing [Fe/H], following the 
entries provided by Harris (1996). The Messier or NGC number of 
the cluster is given in each panel, along with the corresponding 
[Fe/H] value and the telescope employed to obtain the displayed 
data---where ``ESO" stands for the Danish~1.54-m, ``NOT" for the 
Nordic Optical Telescope, and ``HST" for the {\em Hubble Space 
Telescope}. Zero-age HB (ZAHB) models kindly provided by 
VandenBerg et al. (1999), as transformed to the Str\"{o}mgren 
system using Kurucz (1992) color-temperature relations, are also 
shown in each panel; these take into account the $\alpha$-element 
enhancement observed for most metal-poor GCs (e.g., Carney 1996).

In order to fit the ZAHB models to the observations we adopted the
reddenings given in Table~1 (mostly from Harris 1996) and the models 
were then shifted in luminosity until they matched the lower locus 
of the HB stars cooler than the jump. 
For the clusters with data obtained on 
non--photometric nights we have made an effort to adequately fit 
the red end of the HB star distributions; the ZAHB fits for 
clusters with calibrated photometry were used as guidance. Thus we 
have not made use of the reddening values reported in Table 1 for 
these clusters.

Important conclusions can be immediately drawn from an 
inspection of Figure~1:

 1.~The jump in $u$ is a {\em ubiquitous feature}, present 
       in every GC studied which has a sufficiently hot HB.
       (Note that the hottest BHB stars in M30 lie close to the 
       limiting temperature for the occurence of the jump.) 
       Therefore, the effect is by no means restricted to the case 
       of M13, originally investigated by Grundahl et al.\ (1998). 
       Such a jump is morphologically best described as a 
       systematic deviation, in $u$ magnitudes and/or $u-y$ 
       colors, with respect to the expectations of canonical 
       ZAHB models, in the sense that the observations appear 
       brighter and/or hotter than the theoretical predictions;   

 2.~As found by Grundahl et al.\ in the case of M13, {\em the 
       jump occurs at intermediate temperatures only};

 3.~The occurrence of the jump {\em does not depend on 
       metallicity} within the metallicity range of the clusters
       studied here; 

 4.~Both clusters with {\em short} blue tails (e.g., 
       M5,\footnote{The jump in M5 has also been detected by Markov 
       \& Spassova (1999) using broadband ($U$) photometry.} 
       NGC~288) and clusters with {\em extended} blue tails (e.g.,
       M13, NGC~6752) show the $u$-jump, which is thus {\em HB
       morphology-independent}, as long as sufficiently hot 
       BHB stars are present in any given GC; 

 5.~The {\em color $(u-y)_0^{\rm jump}$ appears to change little 
       from cluster to cluster, even when the [Fe/H] values are quite
       different}.

The occurrence of {\em the jump is most decidedly not a 
spurious consequence of the telescope and/or instrumentation 
used}, since it is independently seen with data obtained 
using four filter/detector combinations. We also point out 
that the jump is evident in {\em all instrumental} CMDs as 
well, thus ruling out any problems arising from 
the adopted data reduction or calibration procedures (\S2) as 
a ``cause" for the occurrence of the $u$-jump.

More detailed information can be obtained from the entries in 
Table~1. Before discussing it in depth (\S3.3), we first 
describe our procedure to determine the ``jump parameters" 
$\cal{X}^{\rm jump}$, $(u-y)_0^{\rm jump}$, 
$T_{\rm eff}^{\rm jump}$, and $M^{\rm jump}$.

\subsection{Determining the Low-Temperature ``Cutoff" and 
            the ``Size" of the Jump} 

In order to measure the color $(u-y)_0^{\rm jump}$ which defines 
the onset of the jump at its ``cool" end and estimate the size of
the jump, we have decided to adopt the $\cal{X}$ and $\cal{Y}$ 
coordinates described by Crocker et al. (1988) and 
Rood \& Crocker (1989). As can be seen from Figure~1 in Crocker  
et al., $\cal{Y}$ is indeed ``tailor-made" for measuring the 
departure of HB star distributions from theoretical ZAHBs as seen 
in our CMDs (Fig.~1)---especially since, as already noted, the jump 
is best described as a systematic deviation in $u$ magnitudes 
{\em and/or} $u-y$ colors with respect to the canonical ZAHBs.

%
\begin{figure*}
 \figurenum{1}
 \epsscale{1.750}
 \plotone{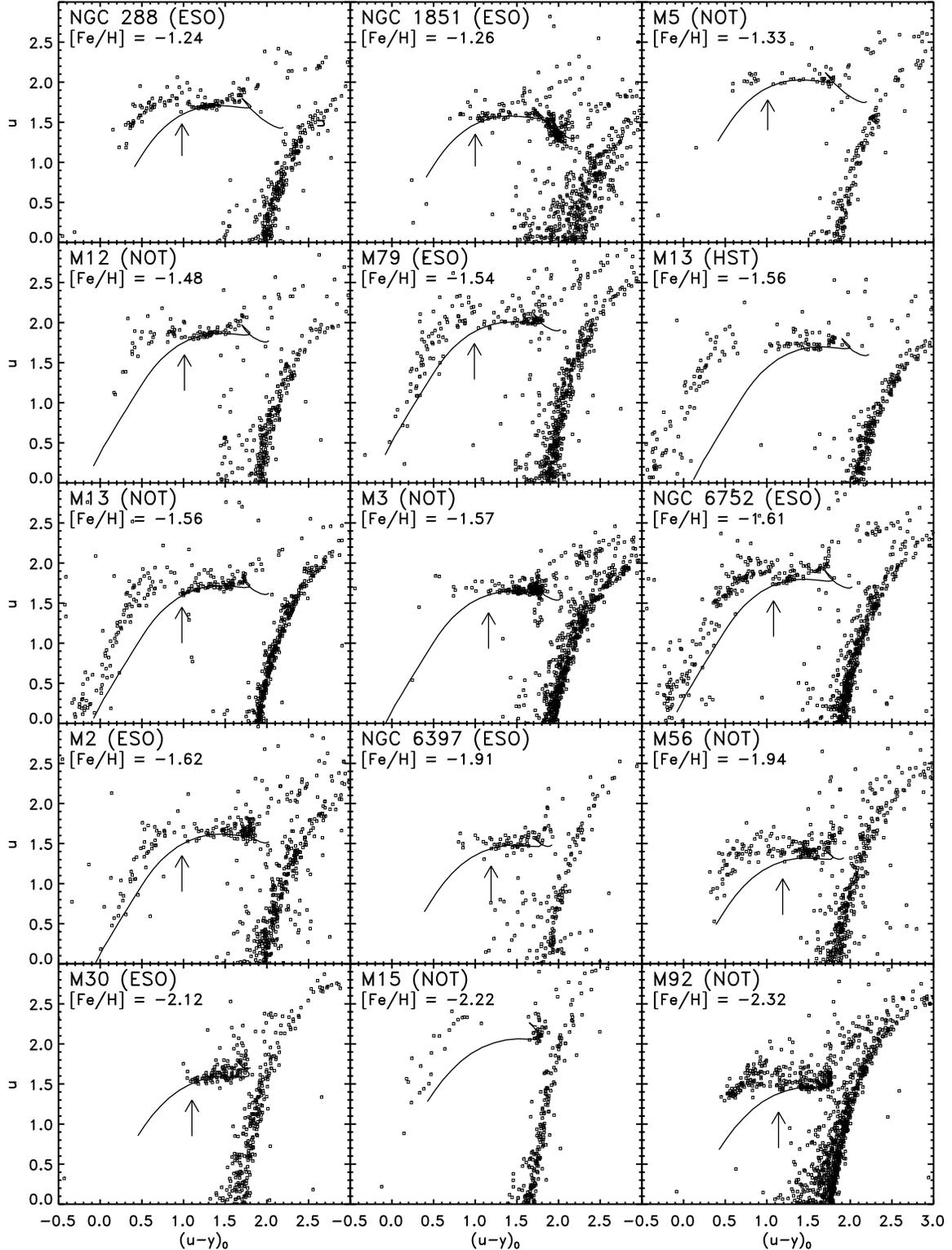}
 \caption{\footnotesize A mosaic plot showing the $u$,~$u-y$ CMDs of 
          fourteen 
          Galactic GCs around the HB region. The cluster names, 
          [Fe/H] values, and an acronym describing the telescope 
          utilized to obtain the corresponding data are shown in 
          each panel, at the upper left-hand corner. The clusters 
          are presented in order of decreasing metallicity from the 
          upper left to the bottom right. The telescope acronyms are 
          as follows. ESO: Danish 1.54-m (Chile); NOT: Nordic Optical 
          Telescope (Canary Islands, Spain); HST: {\em Hubble Space 
          Telescope}. Canonical ZAHB models (from VandenBerg et al.\  
          1999) for the appropriate metallicities are plotted. The  
          required shifts have been applied to account for the reddening  
          and distance modulus of each GC, enforcing satisfactory  
          matches between the data and models at the red end of the 
          distributions. Note that for convenience we have allowed the
          zero point on the luminosity axis to ``float." {\em Note 
          also that a ``jump" at intermediate $u-y$ colors is present 
          in all GCs}, and that its onset, indicated by vertical arrows  
          ($\uparrow$), appears to occur at approximately the same color 
          for them all, irrespective of [Fe/H]. See text for more details.
          The ``glitch'' in the ZAHB near $u-y = 1.6$~mag arises because 
          $u-y$  is not a monotonic function of temperature coolward of 
          the Balmer maximum at $T_{\rm eff} \sim 9000$~K.  
          } 
\end{figure*}

Specifically, for each HB star we measured its projected distance 
from the ZAHB ($\cal{Y}$) as well as the ``path length" ($\cal{X}$)
along the theoretical ZAHB for the appropriate metallicity. The zero 
point for $\cal{X}$ was arbitrarily set at $(u-y)_0\,=\,0.5$~mag. 
In Figure~2,
$\cal{X}$ and $\cal{Y}$ are plotted for all the clusters. $\cal{X}$ 
increases with increasing $u-y$ and $\cal{Y}$ is positive for
stars lying at luminosities higher than the theoretical ZAHB model;
both quantities are measured in magnitudes. 
Since the ZAHB models have been fit to the HB stars cooler than the 
jump, these stars will have $\cal{Y}$ values close to zero.
Dashed horizontal lines are added to locate the $\cal{Y}$~=~0.0 and
$\cal{Y}$~=~0.25 loci. The latter value corresponds to our estimate 
of the change in $\cal{Y}$ (or ``jump size") for M13, as can be 
seen by inspection of the middle left panel in Figure~2. We have not
calculated $\cal{X}$ and $\cal{Y}$ for the HST data set, as we do not
have the models transformed to the appropriate colors and magnitudes. 
The ZAHB overplotted in the HST CMD for M13 refers to the $u$,~$y$ 
system, and not to the WFPC2 filters, and is only intended to guide 
the eye.

%
\begin{figure*}[t]
 \figurenum{2}
 \epsscale{1.55}
 \plotone{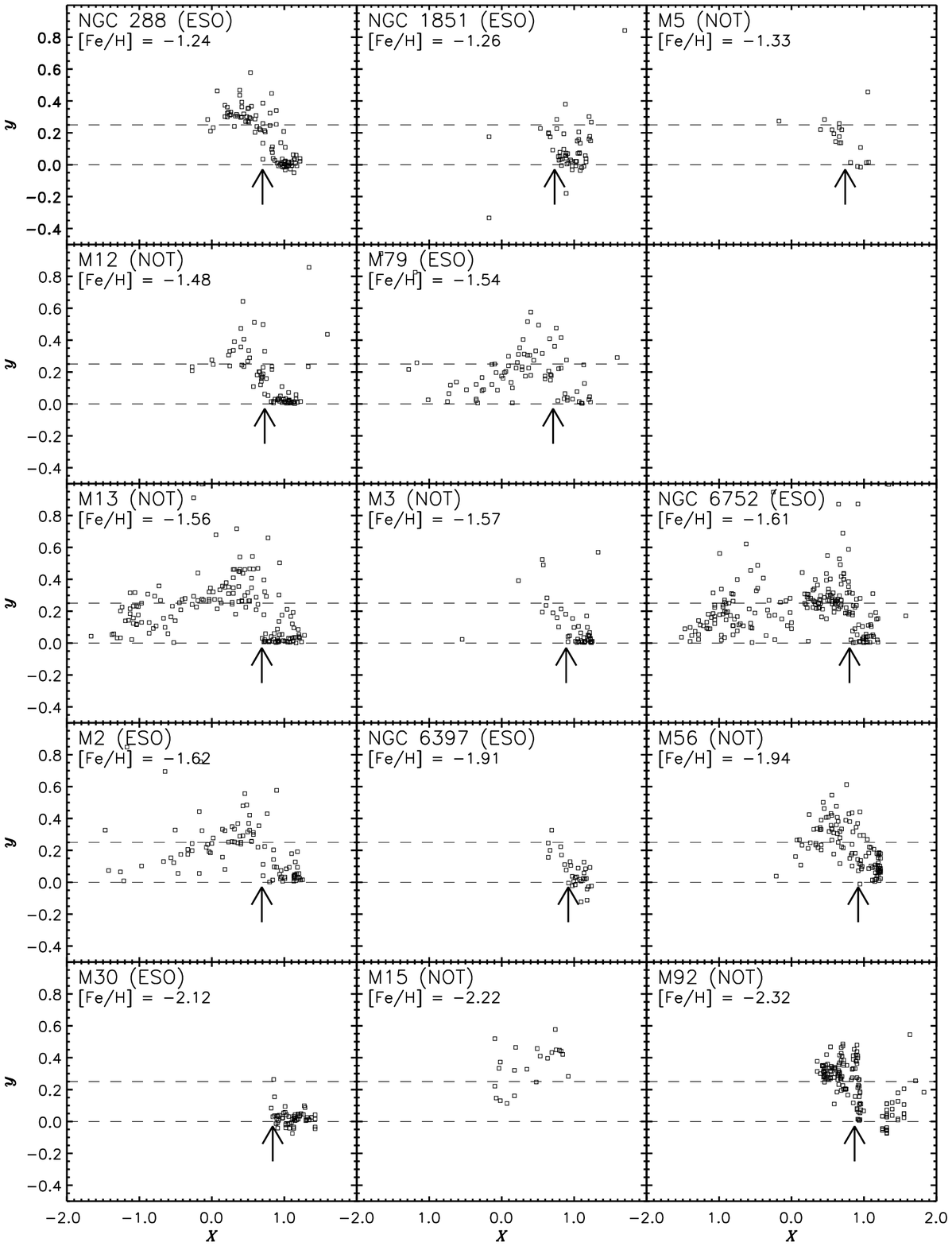}
 \caption{\footnotesize A mosaic plot showing our measured values for 
          $\cal{X}$ and 
          $\cal{Y}$ (\S3.2) for each of the clusters in our sample. The 
          vertical arrows ($\uparrow$) indicate our measured position 
          of the jump, and the dashed lines correspond to  
          ${\cal{Y}}\,=\,0.0$ and ${\cal{Y}}\,=\,0.25$. Note that the 
          reddest HB stars have been omitted from the plot (see \S3.2).
          }     
\end{figure*}

Note that our ZAHB models do not extend to very small envelope 
masses, implying that $\cal{X}$ (and hence $\cal{Y}$) for the 
hottest HB stars cannot be directly estimated on their basis. We 
have therefore extended the ZAHB locus, extrapolating it and adding 
(by hand) an extra point in our $u, u-y$ ZAHB sequences such that 
these stars could be included. Similarly
the detailed morphology of the ZAHB tracks for the coolest HB stars 
leads to some ambiguity in the measurement of $\cal{Y}$ for these 
stars. We have thus excluded these stars from Figure 2. We emphasize 
that our adopted procedure to deal with the hottest/coolest BHB 
stars has {\em no} effect on the conclusions of this paper. 

For all the clusters (Fig.~2)---except M15, for which there 
is a lack of stars---it is easy to determine the 
$\cal{X}$--location of the jump, which we simply estimate by 
eye. In order to assess the error in $\cal{X}^{\rm jump}$ we 
estimated by eye the minimum and maximum ``tolerable" values 
of $\cal{X}^{\rm jump}$ for each cluster and adopted half the 
distance between these two points as our error. Since $\cal{X}$ 
is measured along the theoretical ZAHB it has a one-to-one 
relationship with $(u-y)_0$; we then proceeded to calculate 
$(u-y)_0^{\rm jump}$ from the $\cal{X}$ location of the jump.
An error of 0.01~mag in $u-y$ at the color of the jump 
corresponds to an error of approximately 56~K in temperature. 
The $(u-y)_0^{\rm jump}$ colors thus determined can be found 
in Table~1, along with the estimated errors.\footnote{We also 
estimated $(u-y)_0^{\rm jump}$ directly from the $u, u-y$ 
CMDs presented in Figure~1. In all cases the agreement was 
better than 0.03~mag, which is within the errors in the 
determination of the jump location (see Table~1).} 

Having determined such colors, we evaluated the corresponding 
temperature $T_{\rm eff}^{\rm jump}$ and mass $M^{\rm jump}$ 
values characterizing the canonical ZAHB models for the 
metallicity of the adopted ZAHB model by cubic spline 
interpolation in $(u-y)_0$. These quantities are also given in 
Table~1, and will be discussed in \S3.3 below. Combining the 
error in $(u-y)_0^{\rm jump}$ with the expected photometric 
zero point errors we estimate that the typical errors in 
$T_{\rm eff}^{\rm jump}$ and $M^{\rm jump}$ are 300~K (smaller 
for M13 and NGC~6752) and $0.01\,M_{\sun}$, respectively. Note 
that our error estimates ignore any potential errors in the 
models, as well as uncertainties in the adopted reddening 
values and cluster-to-cluster differences in sample 
size; these may appear as an additional source of random
scatter among the various clusters.

\subsection{The $u$-Jump: a Detailed Empirical Description}  

Table~1 provides detailed quantitative information on the 
nature of the $u$-jump in our set of GCs, as well as some of 
the most relevant physical parameters characterizing the 
latter. The most important implications from this table 
include the following:

 1.~Remarkably, the onset of the $u$-jump occurs at a color 
      $(u-y)_0^{\rm jump}$ which is essentially the same 
      (within the errors) for every GC in our sample (except 
      possibly for M3 and M92), irrespective of metallicity, 
      central density, concentration, or mixing history on 
      the RGB;\footnote{Because we only have one calibrated 
      GC with ${\rm [Fe/H]} < -1.65$ (M92), we caution that a 
      small metallicity dependence of $(u-y)_0^{\rm jump}$ 
      could be present. More data for metal-poor GCs are needed
      to settle this issue. We stress however that {\em if} 
      present such a relation amounts to a change of only 
      $\sim 1000$~K between ${\rm [Fe/H]} = -1.3$ 
      and ${\rm [Fe/H]} = -2.3$.}

 2.~Due to the low dependence of the color-temperature 
      transformations on metallicity, it also follows from the 
      above that the temperature $T_{\rm eff}^{\rm jump}$ is 
      also essentially the same for all GCs in our sample, 
      irrespective of metallicity, central density,
      concentration, or mixing history on the RGB; 

 3.~Unlike $T_{\rm eff}^{\rm jump}$, the mass cutoff 
      $M^{\rm jump}$ is found to depend on metallicity, 
      increasing with decreasing [Fe/H] at a rate 
      ${\rm d} M^{\rm jump}/{\rm d}{\rm [Fe/H]} \approx -0.09\,
      M_{\sun}\,{\rm dex}^{-1}$. Such a mass variation (at an 
      essentially constant temperature) can be ascribed to the 
      behavior of the canonical ZAHB models as a function of 
      metallicity;

 4.~It thus follows that $T_{\rm eff}^{\rm jump}$ is the 
      fundamental quantity characterizing the onset of the 
      $u$-jump, rather than the mass at that point; 

 5.~The size of the $u$-jump is also remarkably constant 
      amongst our sample of GCs, as is evident from inspection 
      of Figure~2;       

 6.~No metal-poor GC is known which does not show a $\log\,g$-jump. 
      Therefore this too seems to be a ubiquitous phenomenon. 
      However, while every GC with a $\log\,g$-jump also 
      shows a $u$-jump, the converse cannot yet be stated with 
      certainty, given that gravities have not yet been measured 
      for an extensive sample of GCs; 

 7.~Importantly, the presence of a $\log\,g$-jump, like that 
      of a $u$-jump, seems to be completely uncorrelated with 
      any physical parameter of the GCs, including the 
      metallicity. From Figure~9 in Moehler et al.\ (1995), one 
      can also see that the boundaries of the $\log\,g$-jump 
      region do not vary as a function of metallicity. Remarkably, 
      {\em the occurrence of the $\log\,g$-jump too does not seem 
      to depend on the mixing history of the GC stars during the RGB 
      phase};

 8.~The presence of the $u$-jump does not depend on the GC 
      dynamics. In our sample, we have loose GCs (M12, M30, 
      NGC~288) showing $u$-jumps which are extremely similar to 
      those found in much more concentrated GCs (M80, NGC~1851). 
      The phenomenon also extends to the realm of core-collapsed 
      GCs with extremely high central densities (M15, NGC~6397, 
      NGC~6752). Again, the same can be said about the $\log\,g$-jump;

 9.~The jump phenomenon---whether $u$- or $\log\,g$- ---does not 
      appear to depend on the distance from the center of the Galaxy. 
      However, since we do not have bulge, disk or outer-halo 
      GCs in our sample, we cannot give support to the more 
      general conclusion that the jump phenomenon does not 
      depend on the stellar population to which the cluster 
      belongs: bulge, disk, inner halo, or outer halo.

Figures~3a ($u$,~$u-y$ plane) and 3b ($\cal{X}$,~$\cal{Y}$ plane)  
show a direct comparison between our calibrated CMDs for  
M13 (circles) and NGC~288 (plus signs). This figure shows that, 
{\em notwithstanding the different metallicities and mixing 
histories on the RGB (see Table~1), M13 and NGC~288 present 
remarkably similar jump location, size, and overall morphology}.

\begin{figure*}
 \figurenum{3}
 \epsscale{1.20}
 \plotone{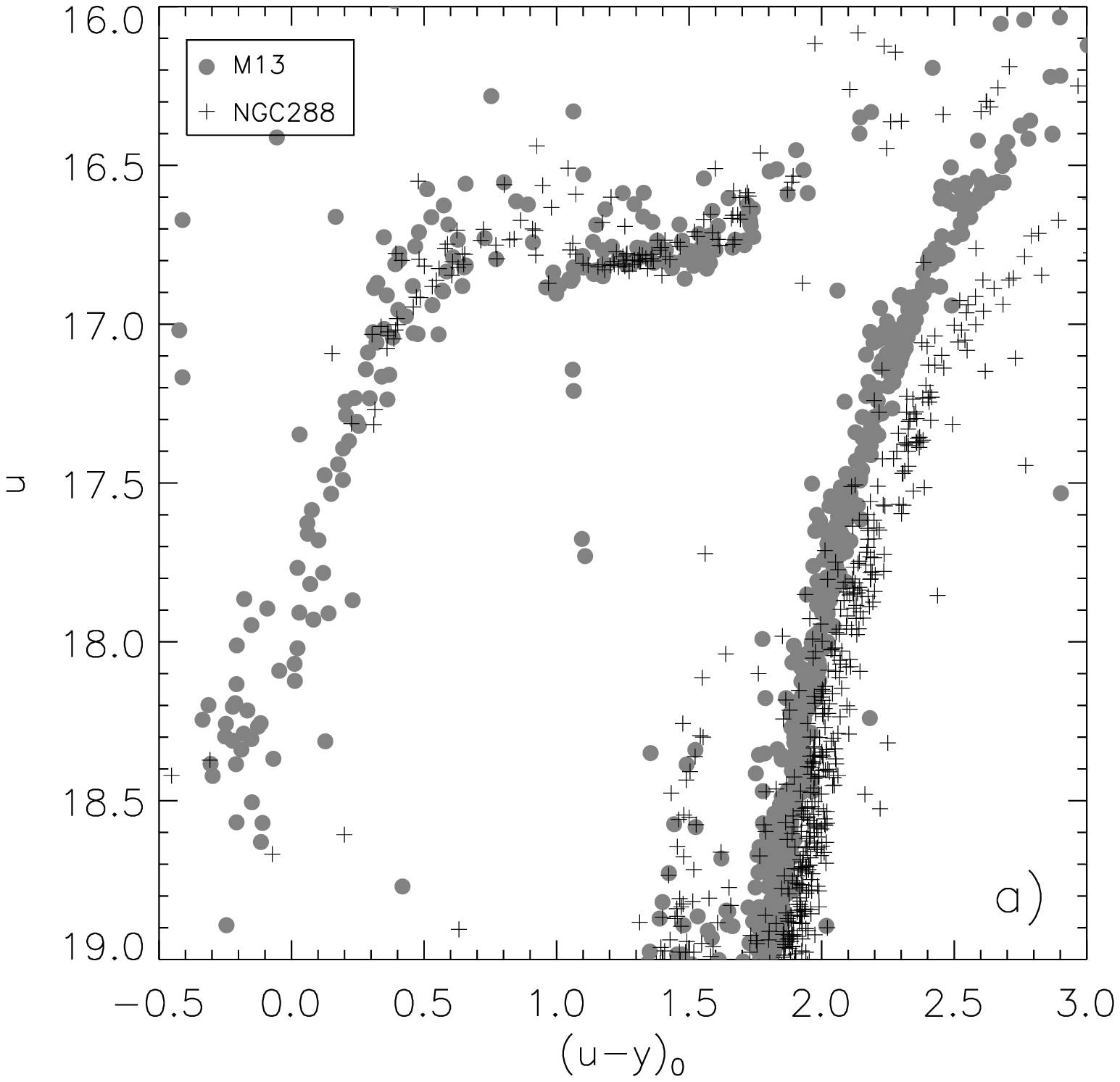}
 \epsscale{1.20}
 \plotone{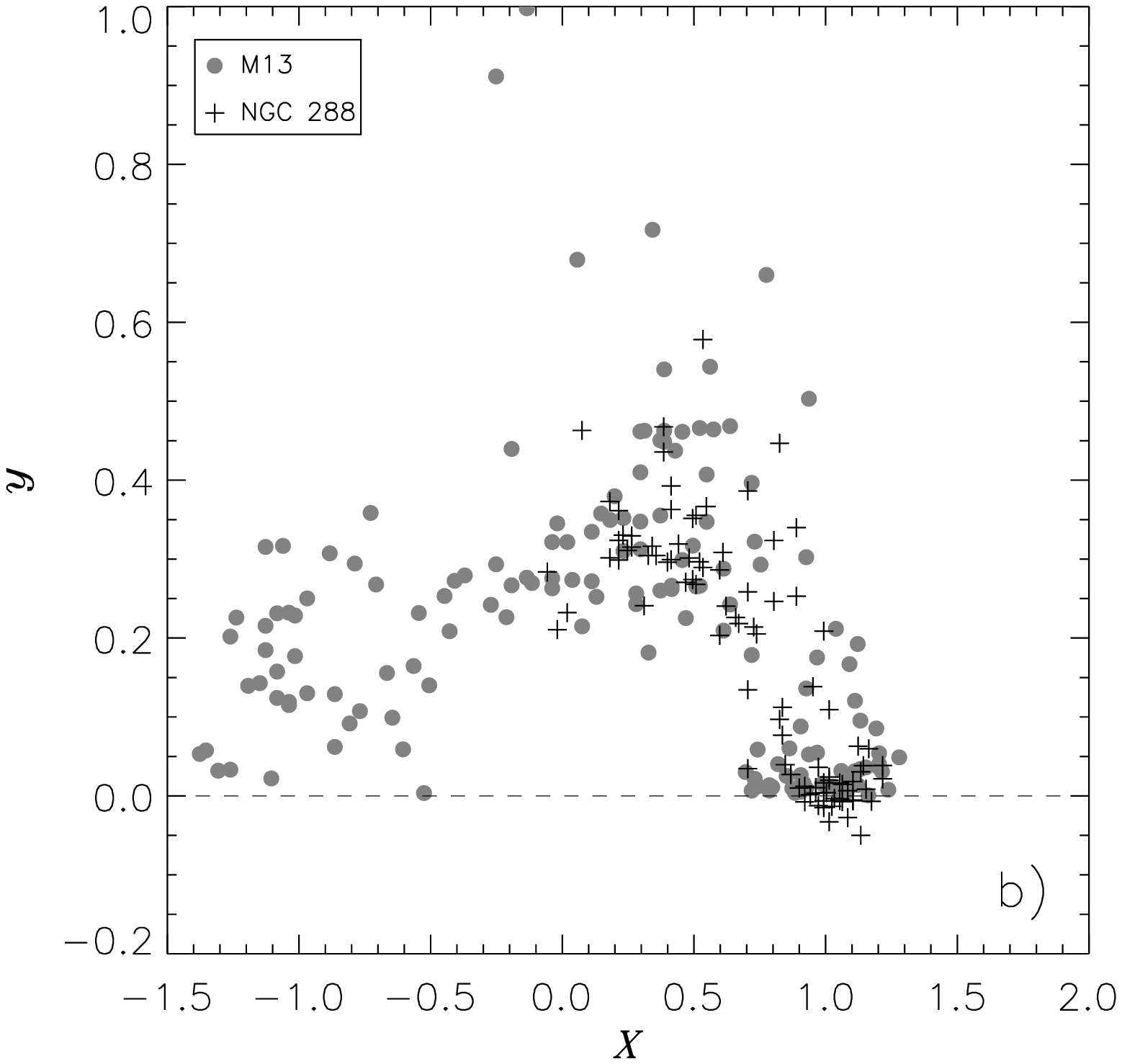}
 \caption{\footnotesize a)~Comparison between the M13 and NGC~288 
          $u$,~$u-y$ CMDs, 
          utilizing the calibrated data for these clusters as described 
          in the text. The symbols ``$\bigcirc$" (in gray) and ``+" (in 
          black) are used for M13 and NGC~288 stars, respectively.  
          Note the metallicity effect upon the RGB color and shape: for   
          M13, ${\rm [Fe/H]} = -1.56$; for NGC~288, ${\rm [Fe/H]} = -1.24$ 
          (Harris 1996). Most importantly, the plot clearly illustrates 
          that both the $(u-y)_0^{\rm jump}$ color and the jump size 
          are essentially 
          the same for the two globulars, notwithstanding the fact that 
          bright M13 RGB stars have definitely undergone very extensive 
          deep mixing, unlike the case in NGC~288 (see Table~1). It 
          thus follows that neither metallicity nor the mixing history 
          on the RGB can be responsible for the ubiquitous nature of the 
          jump among Galactic GCs (Fig.~1). b)~Comparison between M13 and 
          NGC~288 in the $\cal{X}, \cal{Y}$ plane. The symbols have the 
          same meaning as in a).
          }     
\end{figure*}

\subsection{On the Hot End of the Jump} 

In several of the observed clusters it is evident (Figs.~1 and 2) that
stars on the hot side of the jump region again approach the canonical 
ZAHB, as is particularly evident for M13---in which case we estimate a 
temperature of $T_{\rm eff} \sim 20,\!000$~K for the end of the jump 
region. The data presented here for the other clusters with extremely 
long blue tails (M2, M79 and NGC~6752) appear to show that the 
temperature at which stars again approach the ZAHB varies. For these 
three the data were obtained at ESO for the central regions in seeing 
which was significantly poorer than for the M13 observations. 
Consequently we cannot at present decide whether the apparent 
differences for the location of the hot end of the jump are significant 
or due to the effects of poor seeing and crowding. Only observations 
obtained under better seeing conditions can decide this issue.

\section{The Connection between the ``Jump" in Str\"omgren \lowercase{$u$} 
         and Low Blue-HB Gravities}

Analysis of Table~1, as we have seen above, already hints 
that there may be a connection between the $u$-jump and the 
$\log\,g$-jump. We shall now submit this preliminary conclusion 
to a more detailed investigation. 

Figure~4 shows a star-by-star comparison between stars 
which are located inside the $u$-jump region, on the one hand, 
and the $\log\,g$-jump region, on the other hand, for NGC~288 
and NGC~6752 (the two clusters in our sample with the largest 
number of spectroscopic determinations of $\log\,g$ and 
$\log T_{\rm eff}$). Gravities and temperatures were obtained 
from Crocker et al.\ (1988) and Moehler (1999) in the cases of 
NGC~288 and NGC~6752, respectively. As can be seen from this 
figure, stars located in the $u$-jump region (circles) are 
univocally located inside the $\log\,g$-jump region as well. 
Therefore, it is clear that the two effects---the $u$-jump and 
the $\log\,g$-jump---are connected on a star-by-star basis.

%
\begin{figure*}[t]
 \figurenum{4}
 \epsscale{0.40}
 \plotfiddle{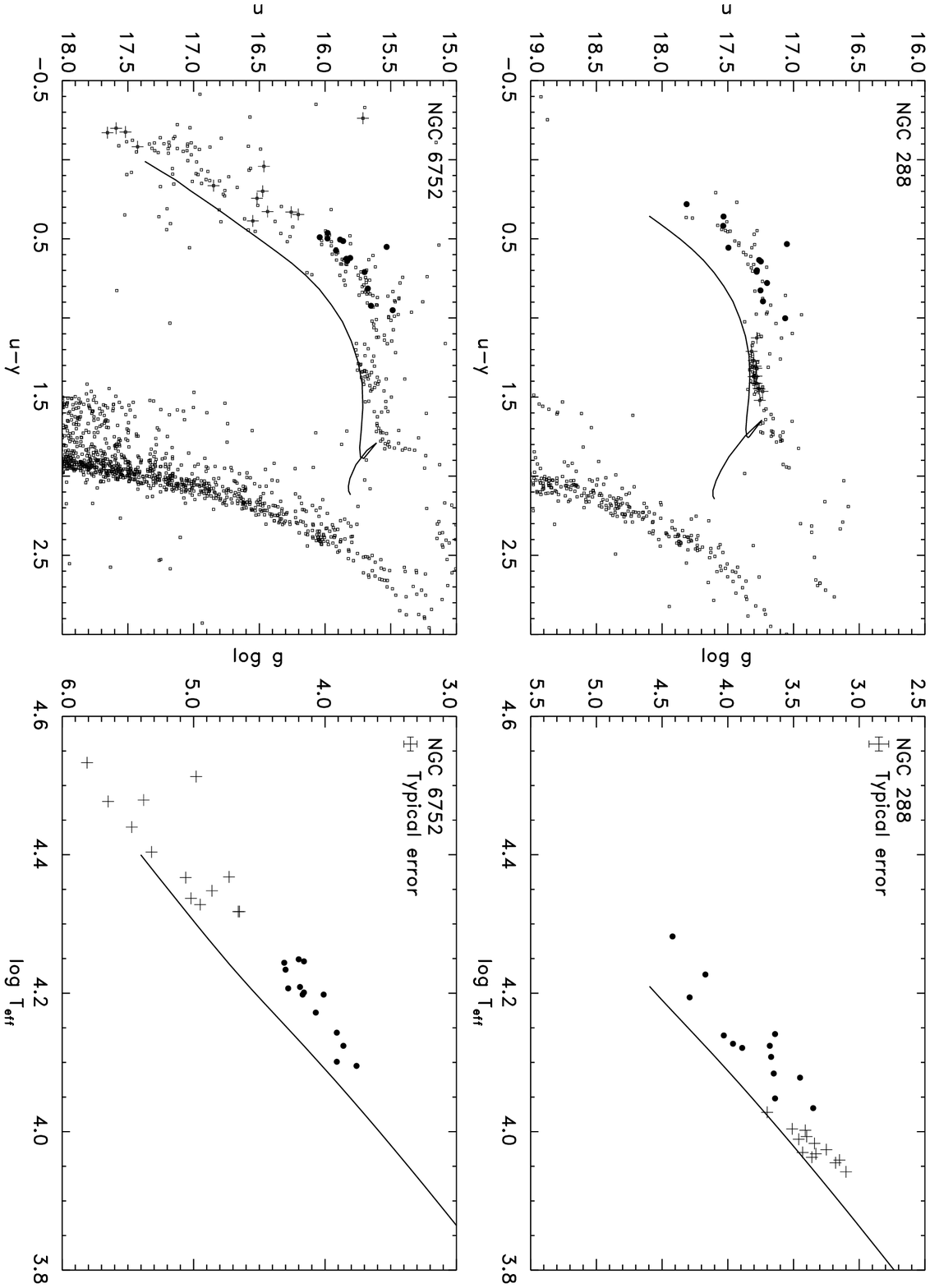}{5.25in}{90}{75}{75}{295}{-36}
 \caption{\footnotesize Cross-check between the location of NGC~288 
          and NGC~6752 
          BHB stars on the $u$,~$u-y$ plane (left column) and 
          on the $\log\,g$,~$\log\,T_{\rm eff}$ plane (right column). 
          The small open squares ($\Box$) represent our photometry for 
          the cluster stars. Stars located in the jump region which have
          spectroscopically determined $\log\,g$ and $\log\,T_{\rm eff}$
          are overplotted as small filled circles ($\bullet$), whereas 
          stars located outside the jump region with spectroscopic 
          measurements are overplotted as plusses (+). Note that stars 
          classified as jump stars based on their location in the 
          $u$,~$u-y$ plane (left panels) are also seen to be located 
          in the gravity jump region thus demonstrating that the 
          ``$u$-jump" and the ``$\log\,g$-jump" are closely connected.
          }     
\end{figure*}

This result is also evident from an analysis of Figure~5. This plot 
shows the $\log\,g$, $\log\,T_{\rm eff}$ diagram for stars which 
have had their positions in the $u$, $u-y$ diagrams evaluated on 
the basis of our photometry for several different GCs. Stars which 
are found to lie {\em inside} the $u$-jump region are plotted with 
black symbols, whereas those lying {\em outside} the $u$-jump 
region are shown with gray symbols. It is clear that the vast 
majority of the stars investigated conform to the notion that 
{\em the $u$-jump and the $\log\,g$-jump are different 
manifestations of one and the same physical phenomenon}. The few 
stars which appear not to follow the rule---located exclusively 
at either the very hot or very cool ends of the jump region---can 
easily be accounted for on the basis of observational errors.

%
\begin{figure*}[b]
 \figurenum{5}
 \epsscale{1.60}
 \plotone{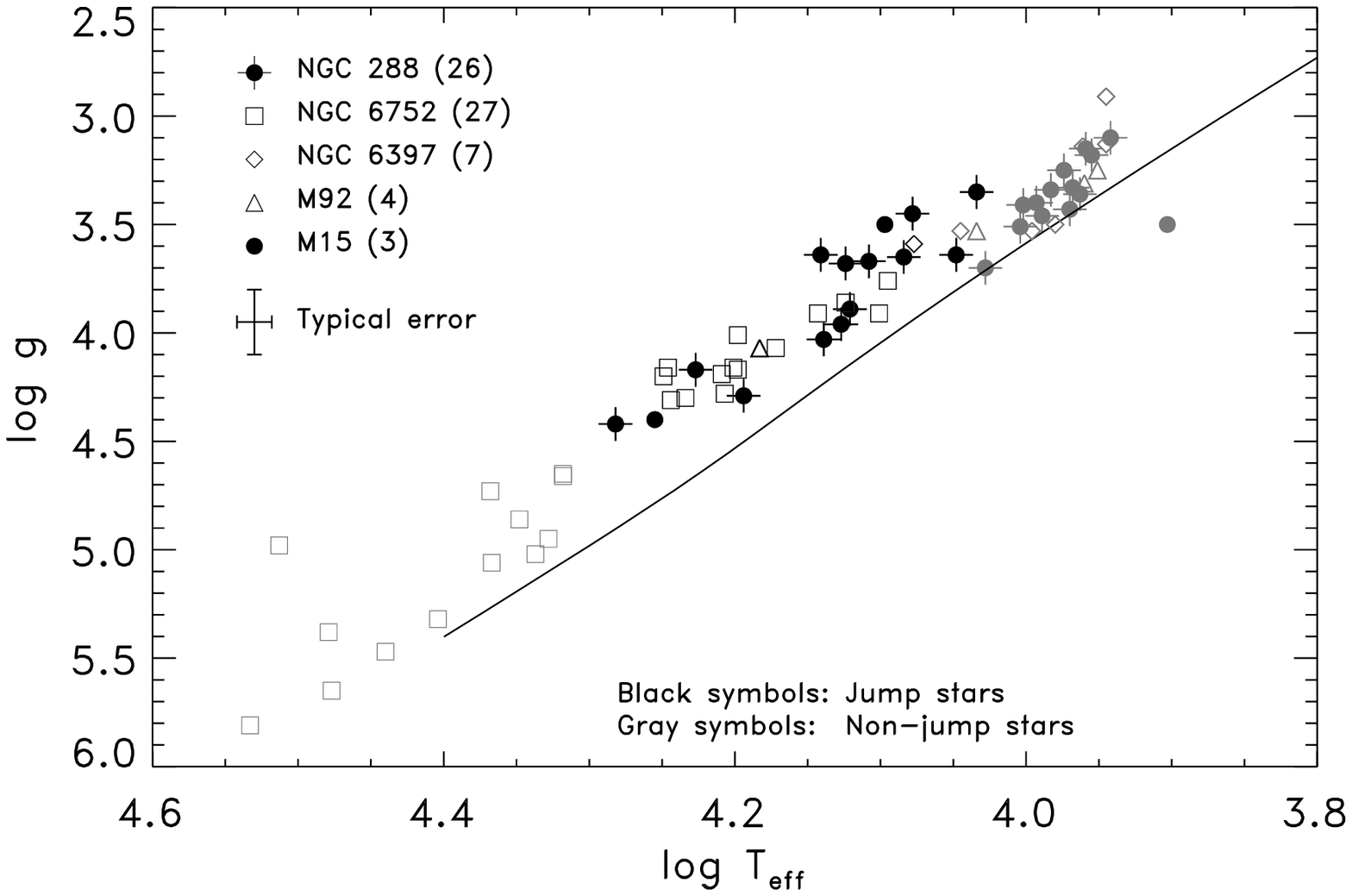}
 \caption{\footnotesize Graphical demonstration of the close connection 
          between 
          the so-called ``low-gravity stars" and the $u$-jump stars. 
          This $\log\,g$,~$\log\,T_{\rm eff}$ diagram shows exclusively 
          stars which have had their positions determined also in 
          the $u$,~$u-y$ plane. (For an impressive diagram showing most 
          of the stars with published gravities to date, see Fig.~9 in 
          Moehler et al.\ 1995.) The number of stars for which both 
          ($\log\,g$,~$\log\,T_{\rm eff}$) and ($u$,~$u-y$) data are 
          available is given in parentheses next to the name of the 
          cluster (upper left-hand corner of the figure). 
          {\em Stars which lie inside the $u$-jump region (in the CMD)
          are plotted as black symbols, whereas stars which lie outside 
          this region are plotted with gray symbols.} 
          }     
\end{figure*}

\section{Constraints on Helium Mixing}

As an explanation for the $u$-jump, Grundahl et al.\ (1998) tentatively 
suggested that very deep mixing during the RGB phase---reaching, in fact, 
all the way into the hydrogen-burning shell and leading to non-canonical 
dredge-up of helium to the envelope---could provide an explanation for 
their observations. This would appear to be an especially compelling 
explanation in the case of M13, for which deep mixing among RGB stars 
is extremely well documented (see Table~1).  

Such helium mixing was first conjectured by VandenBerg \& Smith (1988),  
who highlighted the implications it would have upon our understanding of 
such problems involving the HB phase as the period-shift effect (\S1). 
The idea was later revived by Langer \& Hoffmann (1995), and especially  
by Sweigart (1997a, 1997b). It is generally assumed that mixing processes 
on the RGB are somehow related to stellar rotation, as in the meridional 
circulation theory (Sweigart \& Mengel 1979; see also Kraft 1994, 1998, 
1999 and Sneden 1999 for recent reviews).

As pointed out by Sweigart (1997b), one key aspect of the helium mixing 
theory is that Al enhancements, according to RGB nucleosynthesis models 
computed by Langer, Hoffman, \& Sneden (1993), Langer \& Hoffman (1995), 
Cavallo, Sweigart, \& Bell (1996, 1998), etc., can only be produced 
{\em inside} the hydrogen-burning shell.\footnote
{     
    As discussed by Kraft (1998, his \S4.1), such model predictions are 
    at odds with the available determinations of the Mg-isotope ratios 
    among bright GC giants which suggest that Al is produced at the 
    expense of $^{24}$Mg (e.g., Cavallo 1997). As 
    emphasized by Shetrone (1996a, 1998b) and others, ``using the 
    current nuclear cross-sections $^{24}$Mg can be converted into 
    Al but only at temperatures higher than those found in the CNO
    [hydrogen-burning] shell!" (Shetrone 1998b). In fact, such 
    temperatures should be {\em substantially} higher than that 
    found at the H-burning shell in the models (e.g., Langer, 
    Hoffman, \& Zaidins 1997; Denissenkov et al.\ 1998): 
    $\approx 70 \times 10^{6}$~K, as opposed to  
    $\simeq  55 \times 10^{6}$~K. Mixing to such high temperatures is 
    completely ruled out by canonical evolutionary theory, and is not 
    envisaged in Sweigart's (1997a, 1997b) helium-mixing theory either. 
    Thus, the basic nuclear-reaction mechanism which lies at the 
    root of the helium-mixing scenario remains unsettled. And, as 
    emphasized by Denissenkov et al.\ (1998), ``~`unfortunately' 
    [sic], nuclear physicists seem to have little (if any) doubt 
    concerning the current $^{24}$Mg(p,$\gamma$)$^{25}$Al reaction
    rate". For further discussion, the reader is referred to the 
    interesting papers by Smith \& Kraft (1996), Langer et al.\ 
    (1997), and Denissenkov et al.\ (1998). 
} 
Hence, any Al overabundance should 
necessarily be accompanied by the dredge-up of helium freshly 
produced inside the shell. This is a particularly important result, 
given that large Al overabundances are indeed observed among RGB 
stars in several GCs [see Table~1, and also Norris \& Da Costa 
1995a, 1995b and Zucker, Wallerstein, \& Brown 1996 for the impressive 
case of $\omega$~Cen (NGC~5139); recent reviews have been provided by 
Kraft 1994, 1998, 1999 and Sneden 1999]. If helium mixing were 
present among Galactic GCs, one would expect a correlation between HB 
morphology and O, Na, Mg, and Al abundance variations on the RGB. 
Indeed, a correlation between HB morphology and the presence/extent of 
signatures of deep mixing on the RGB has been independently suggested 
by several different authors (Catelan \& de Freitas Pacheco 1995; Kraft 
et al.\ 1995, 1998; Peterson, Rood, \& Crocker 1995; Carretta \& Gratton 
1996). In fact, helium mixing stands out as the best candidate to 
explain the anomalous HB morphology of the ``metal-rich" GCs NGC~6388 
and NGC~6441 (Sweigart \& Catelan 1998; Layden et al.\ 1999). 

We thus attempt to ascertain the extent to which helium mixing on the 
RGB may be responsible for the $u$- and $\log\,g$-jump phenomenon.

\subsection{Constraints from the Morphology of the ($u$,~$u-y$) and
($\log\,g$,~$\log\,T_{\rm eff}$) Diagrams} 

Sweigart (1997b) has shown that it is possible to reproduce the 
$\log\,g$, $\log\,T_{\rm eff}$ pattern seen among all GC BHB 
stars observed to date by invoking helium mixing on the RGB. It is 
useful to recall what requirements such helium-mixed stars would 
have to fulfill in order to explain the jump phenomenon. 

Expanding on Sweigart's (1997b) scenario, one would expect the 
following behavior as a function of ZAHB temperature:

 1.~$\log\,T_{\rm eff} \lesssim 4.0$: HB progenitors (i.e., RGB 
      stars) do not experience significant helium mixing, and HB
      stars accordingly lie along canonical evolutionary tracks; 

 2.~$4.0 \lesssim \log\,T_{\rm eff} \lesssim 4.2$: HB stars are 
      somewhat more luminous than canonical models due to a larger 
      helium abundance ($Y \sim 0.30 - 0.35$) in their envelopes;

 3.~$4.2 \lesssim \log\,T_{\rm eff} \lesssim 4.3$: The increase 
      in envelope helium abundance due to deep mixing becomes very 
      large, $Y \sim 0.40 - 0.45$;

 4.~$\log\,T_{\rm eff} \gtrsim 4.3$: The HB luminosity is 
      dominated by the helium-burning core (as opposed to the 
      hydrogen-burning shell at lower temperatures, which is now 
      inert), and so the helium-mixed and canonical models 
      essentially agree---even though the envelope helium abundance 
      in the helium-mixed models can be $Y \gtrsim 0.45$.

Unfortunately, it is not possible to directly 
test for the presence of enhanced surface helium because 
helium is generally observed to be depleted in the photospheres 
of hot HB stars (e.g., Moehler et al.\ 1997), most likely due to 
diffusion processes (\S6).  

We suggest, however, that the helium mixing pattern among BHB 
stars, as described above, is unlikely. We have four main 
arguments against helium mixing as an explanation for the jump
based on the morphology of the $u$,~$u-y$ and
$\log\,g$,~$\log\,T_{\rm eff}$ diagrams:

 1.~The variation of $Y$ with temperature required by Sweigart 
      (1997b), not having been derived from first principles,  
      can only be achieved by fine-tuning of free parameters in the 
      helium-mixing theory. Even 
      if one assumes that there is a ``cutoff rotational velocity" 
      beyond which mixing occurs, but below which no mixing 
      takes place, it seems virtually impossible to 
      produce a low-temperature cutoff for the jump which is so 
      remarkably constant---i.e., to within $\pm \approx 
      500$~K---from one GC to the next, given the strong dependence 
      of ZAHB properties upon variations in GC evolutionary 
      parameters (e.g., Sweigart \& Gross 1976). In fact, given 
      that the HB effective temperature becomes less sensitive to 
      changes in mass as the metallicity decreases (see Fig.~7 in 
      Buonanno, Corsi, \& Fusi Pecci 1985), one would expect some 
      intrinsic relationship between $T_{\rm eff}^{\rm jump}$ and 
      [Fe/H]. As already mentioned (\S3.3), any intrinsic 
      relationship between $T_{\rm eff}^{\rm jump}$ and [Fe/H],  
      if present at all, seems to be quite mild. 
      This may be called the {\em ``global" fine-tuning problem}. 
      This global fine-tuning problem is a major impediment facing 
      {\em any} stellar evolution-related scenario for the jump, 
      probably pointing instead to a stellar atmospheres-based 
      explanation (\S6);  

 2.~Fine tuning 
      is also required in the helium mixing scenario 
      {\em at any given metallicity} and {\em for any given GC}. 
      Quantitative information in this respect can be obtained 
      from detailed inspection of the plots published by Sweigart 
      (1997a, 1997b). Figure~4 in Sweigart (1997a) is particularly 
      relevant in this regard. This figure shows how the expected 
      ZAHB temperature increases with increasing values of both 
      Reimers' (1975a, 1975b) mass loss parameter, $\eta_{\rm R}$, 
      and the deep mixing depth, $\Delta X$. Thus in order 
      to produce a jump at fixed $T_{\rm eff}$ an increase in 
      the mixing extent must be compensated by a decrease in the 
      mass loss parameter. At $\log\,T_{\rm eff} = 4.1$, which is 
      very close to the empirical value for $T_{\rm eff}^{\rm jump}$ 
      (see Table~1), the following combinations 
      ($\Delta X$,~$\eta_{\rm R}$) are found:  
      (0.00,~0.52); (0.05,~0.46); (0.10,~0.40); (0.20,~0.30). 
      If we relax the fixed-$T_{\rm eff}$ 
      constraint and keep instead not only the age, metallicity and 
      original helium abundance, but also $\eta_{\rm R}$ 
      fixed (which is the more natural assumption), 
      a {\em very} large gap in $\log\,T_{\rm eff}$ results.  
      Figures 7 through 9 in Sweigart (1997b) show how the 
      gravity-temperature plane is affected by the extent of helium 
      mixing on the RGB. From those plots one can infer that the 
      substantial increase in $Y$ that would be required to 
      reproduce the observed jump would lead to a gap in 
      temperature encompassing several thousand degrees Kelvin, 
      besides leading to an increase in gravity (and hence 
      a {\em decrease} in luminosity). If some natural variation  
      in $\eta_{\rm R}$ is invoked, one would most likely 
      expect---due to the widely suggested mixing-rotation 
      connection---that $\eta_{\rm R}$ should actually 
      {\em increase}, and not decrease, with increasing 
      mixing extent (see, e.g., \S6.2 in Kraft et al.\ 1995),  
      as opposed to what would be required to eliminate the gap. 
      These perhaps surprising predictions of the 
      helium-mixing scenario have no counterpart in either the 
      $u$,~$u-y$ or the $\log\,g$, $\log\,T_{\rm eff}$ 
      diagrams, and demonstrate the high degree of fine tuning 
      required for helium mixing to account for the jump phenomenon 
      at a given metallicity and for any given GC. This may be 
      called the {\em ``local" fine-tuning problem}---not to be 
      confused with the global, metallicity-related fine-tuning 
      problem described above; 

 3.~Without appealing to ad-hoc hypotheses, the jump location  
      {\em and} size should depend quite strongly (in the helium 
      mixing scenario) on the extent of deep mixing on the RGB. 
      However, GCs in which the RGB stars have undergone extreme 
      deep mixing---such as M13---present jump characteristics 
      virtually identical to those of GCs whose giants seem to 
      have undergone much less extensive deep mixing---such as 
      NGC~288 (see Figs.~1, 3, and 5, and also Table~1);

 4.~In a similar vein, if deep mixing is related to stellar 
      rotational velocity, and given that there is no {\em a priori} 
      reason to expect rotational velocities at a given metallicity 
      to be the same from one cluster to the next (as supported by 
      the observations of, e.g., Peterson, Rood, \& Crocker 1995
      and references therein), one would definitely expect large 
      intrinsic scatter in $(u-y)_0^{\rm jump}$, 
      $T_{\rm eff}^{\rm jump}$, and jump size at any given 
      metallicity unless one resorts to ad-hoc hypotheses. 
      While there {\em does} seem to be a perceptible difference 
      in the jump temperature for M3 and M13 (Table~1)---which 
      might perhaps be related to the difference in HB rotational 
      velocities between the two (see \S6)---we note that the jump 
      {\em size} appears very similar for all clusters (Fig.~2). 

We shall present an alternative scenario to explain the jump 
phenomenon in \S6 below.

\subsection{The Role of Field Stars} 

It is well known that RGB stars in the field do not show (deep) 
mixing patterns nearly as large as GC giants. Since Kraft et al.\ 
(1982), the literature has become very extensive in this regard: 
e.g., Sneden et al.\ (1991, 1997); Kraft et al.\ (1992); Pilachowski 
et al.\ (1996); Shetrone (1996b); Hanson et al.\ (1998); Kraft (1994, 
1998, 1999); Carretta et al.\ (1999b); etc. If helium mixing is 
responsible for the jump, one would reach the conclusion that 
{\em field BHB stars should not show any evidence for a jump in $u$ 
or $\log\,g$} similar to that seen amongst GCs. 

However, as can be seen from the $\log\,g$,~$\log\,T_{\rm eff}$ 
diagrams obtained by Saffer et al.\ (1994, 1997), and most recently 
by Mitchell et al.\ (1998, their Fig.~5), cluster and field BHB 
stars are clearly closely related as far as the jump morphology goes.  
In fact, according to Saffer (1998) ``the cluster and field BHB 
distributions in the $\log\,g$,~$\log\,T_{\rm eff}$ plane are 
completely consistent with one another." This implies that deep 
mixing is unlikely to be the primary cause for the jump phenomenon. 

What is the evidence for a $u$--jump among the field BHB stars? In 
order to answer this question one should ideally have a sample
of BHB stars with 
accurately determined distances and well calibrated Str{\"o}mgren 
photometry, such that their absolute magnitudes could be reliably 
derived and plotted in a $[(u-y)_0, M_u]$ diagram as for the 
clusters. However to the best of our knowledge a sample of BHB 
stars with accurately determined distances does not currently 
exist in the literature. Thus at present we cannot shed further 
light on the connection between the $u$- and $\log\,g$-jumps 
for field BHB stars. 

Whereas the {\em overall} field HB population is believed to contain 
only a small fraction of EHB stars ($\lesssim 1\%$; Saffer \& Liebert 
1995; Villeneuve et al. 1995), the {\em halo} field appears to 
contain a surprisingly large population of sdB (or EHB) stars, 
if compared to the disk field. Mitchell (1998) estimates that 
``the metal-poor halo population can produce a horizontal-branch 
morphology that is, by a factor of $\gtrsim 7$ [$2\sigma$ lower 
limit], more heavily weighted toward the `extreme' blue end than 
the horizontal branch produced by the relatively metal-rich disk 
population". Assuming Mitchell's arguments to be correct, this, 
along with the lack of abundance anomalies among field metal-poor 
giants, could imply that most halo EHB stars do not originate from 
deep mixing processes on the RGB evolutionary phase. This, of 
course, would {\em not} rule out the possibility that some EHB 
stars in some GCs---especially, of course, those showing extreme 
mixing patterns on the RGB---may indeed have undergone helium 
mixing during the RGB phase. More work is needed to verify 
Mitchell's results.

\subsection{Constraints from the Ultraviolet Photometry of GCs}

Is the $u$-jump reported in this paper due to a deviation in the 
bolometric luminosity from canonical HB models, or is it due to a 
spectral peculiarity which makes the $u$ band brighter without 
changing the bolometric luminosity? In principle, ultraviolet 
photometry can be used to answer this question because the stars 
hotter than the jump temperature emit most of their bolometric 
luminosity in the ultraviolet. For example, using the model 
atmosphere tabulation of Lejeune, Cuisinier, \& Buser (1997), one 
finds that a star  with $T_{\rm eff} = 16,\!000$~K, $\log\,g =4.0$, 
and ${\rm [Fe/H]} = -1.5$ emits 73\% of its bolometric luminosity 
shortward of 3000~\AA.  

Ultraviolet photometry of GCs has been obtained using both the
Ultraviolet Imaging Telescope (Stecher et al.\ 1997) and the 
ultraviolet (F160BW, F218W, F255W) filters on WFPC2 (e.g., Sosin 
et al.\ 1997; Ferraro et al.\ 1998). The instruments are 
complementary in that the UIT had a large ($40'$ diameter) field 
of view, but a relatively coarse ($3''$) spatial resolution which 
made it mainly useful in the outer regions of the clusters, whereas 
the WFPC2 images have much better resolution (0\farcs1) but can 
only record significant HB number counts in the cluster cores, due 
to its much smaller field of view. We note that the comparison of 
ultraviolet CMDs with absolute theoretical luminosities can be 
problematic because the reddening correction is large, and the 
ultraviolet reddening law is known to show spatial variations in 
the Galaxy (Fitzpatrick 1999). 
In addition, UIT had a calibration anomaly reminiscent of (but not 
identical to) reciprocity failure (Stecher et al.), while 
the absolute photometry using the Wood's (F160BW) filter is limited
by a high contamination rate (Whitmore, Heyer, \& Baggett 1996) 
and a PSF that varies across the field (Watson et al.\ 1994). 
However, these absolute calibration difficulties are not important 
when  looking for an analog of the \stromu\ jump in the ultraviolet.  
We shall assume that the absolute level of the model ZAHB has been 
adjusted to match the cooler ($T_{\rm eff} < 10,\!000$~K) HB 
stars, and look for an offset from the ZAHB for the hotter stars.

The most accurate GC photometry from UIT was obtained for 
NGC~6752 (Landsman et al.\ 1996). Not only did NGC~6752 have 
the deepest UIT exposure of any globular, but the cluster is 
sufficiently nearby that IUE spectra of 14 hot HB stars are 
available to verify the calibration. 
For $\log\,T_{\rm eff}<4.3$, the (1620~\AA) ultraviolet CMD 
of Landsman et al.\ shows excellent agreement with the canonical 
HB tracks of Sweigart; at higher temperatures the data fall 
$0.1-0.2$~mag below the models (also see Fig.~8). A very
similar result is found for the UIT photometry of M79 by Hill et al.\ 
(1996). In the UIT CMD of $\omega$~Cen (Whitney et al.\ 1998), there 
is a significant population of stars hotter than $16,\!000$~K lying 
{\em below} the (${\rm [Fe/H]} = -1.5$) ZAHB, but again there is no 
evidence for a photometric jump corresponding to that observed in 
$u$. Only for the case of M13 do the UIT data show a possible 
analog of the $u$-jump. Parise et al.\ (1998) report an offset 
toward higher luminosity for stars with 
$4.1 < \log\,T_{\rm eff} < 4.3$.     

Turning to  ultraviolet WFPC2 data, Sosin et al.\ (1997) find 
an excellent fit of ZAHB models to the (F218W,~$B$) CMD of 
NGC~2808, after an adjustment of the zero-point calibration. 
Rood et al.\ (1998) show a good model fit to both the (F160BW,~$V$) 
and (F255W,~$V$) CMDs for M13, and their data suggest a similar 
result for M80. Because the WFPC2 result on M13 is in apparent 
contradiction to the UIT result of Parise et al.\ (1998), we have 
carried out a more detailed examination of both sets of data. We 
have performed our own reduction of the UIT data, while we have 
used the WFPC2 photometry kindly supplied by Ferraro \& Paltrinieri 
(1999). Both CMDs are shown in Figure~6. Evidently, there is a 
problem in the absolute calibration in one or both data sets, because 
there is a 0.25~mag difference in the distance modulus needed to match 
a theoretical ZAHB to the cooler stars. However, the overall appearances 
of the CMDs are quite similar to each other, and to the ultraviolet CMD 
of NGC~6752 shown in Figure~8. The most striking difference is that 
the number ratio of cool to hot HB stars is higher for the UIT data, 
possibly suggesting a radial gradient in HB morphology, with the HB 
morphology being bluer in the core. In the ultraviolet CMDs of both M13 
and NGC~6752, the data fall 0.1 to 0.2 mag below the models at the highest 
temperatures. (Rood et al. 1998 do not find this discrepancy, apparently 
because of their use of the oxygen-enhanced HB models
of Dorman et al.\ 1995.)
The offset to higher luminosity 
reported by Parise et al.\ is present in the UIT CMD for stars with 
$-3.4 < m_{162}-V < -2.1$ 
[$13,\!600~{\rm K} < T_{\rm eff} < 21,\!100$~K], 
and present to a lesser extent in the WFPC2 data. Note that this offset 
occurs at a $m_{162}-V$ color which is 0.75~mag bluer than would be 
predicted from the temperature ($\log T_{\rm eff} = 4.07$) of the jump 
found in the \stromu\ CMD. Also note that this offset in the ultraviolet 
CMD is best described as an absence of stars near the ZAHB, since the 
majority of the stars are still contained within the same empirical upper 
envelope that fits stars at lower and higher temperatures---{\em unlike 
the case with the $u$-jump and the $\log\,g$-jump}.    Thus, while 
there is some evidence for a luminosity offset in the ultraviolet 
CMD of M13, it does not appear to be simply connected to the jump 
observed in the \stromu\  CMD.

%
\begin{figure*}  
 \figurenum{6}
 \epsscale{1.750}
 \plotone{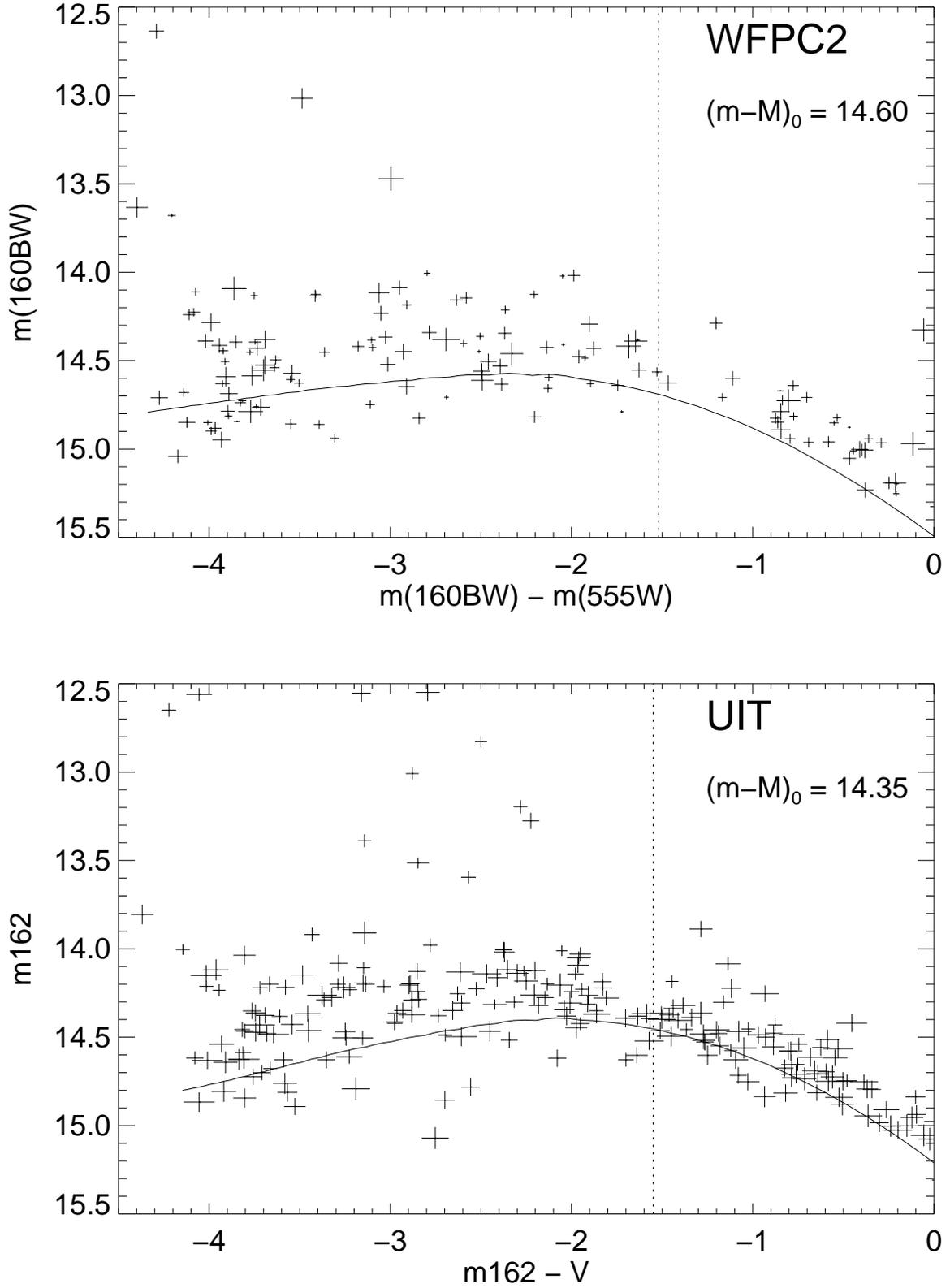}
 \caption{\footnotesize Ultraviolet CMDs of M13 obtained using WFPC2 
          imaging
          of the cluster core with the F160BW filter (upper panel), 
          and using UIT ($\sim 1620$~\AA) imaging of the outer regions 
          of the cluster (lower panel). For each CMD, a canonical 
          ZAHB from Sweigart (see Landsman et al. 1996) with 
          ${\rm [Fe/H]} = -1.6$ which fits the cooler HB stars has been 
          overplotted; note the different adopted distances in the two 
          figures. The vertical dotted line on each CMD indicates the 
          color corresponding to the ``jump" temperature 
          ($\log\,T_{\rm eff}$ = 4.07) found for M13 from the 
          Str\"{o}mgren $u$ CMD. The slightly different shape of the
          ZAHB in the two panels is due to the fact that, although the 
          WFPC2 F160BW filter and the UIT 1620~\AA\ 
          (B5) filter both have effective wavelengths near 1600~\AA, 
          the width of the F160BW filter is approximately twice that of 
          the UIT filter ($\Delta \lambda \sim 225$ \AA).  
          } 
\end{figure*}

%
\begin{figure*}
 \figurenum{7}
 \epsscale{1.750}
 \plotone{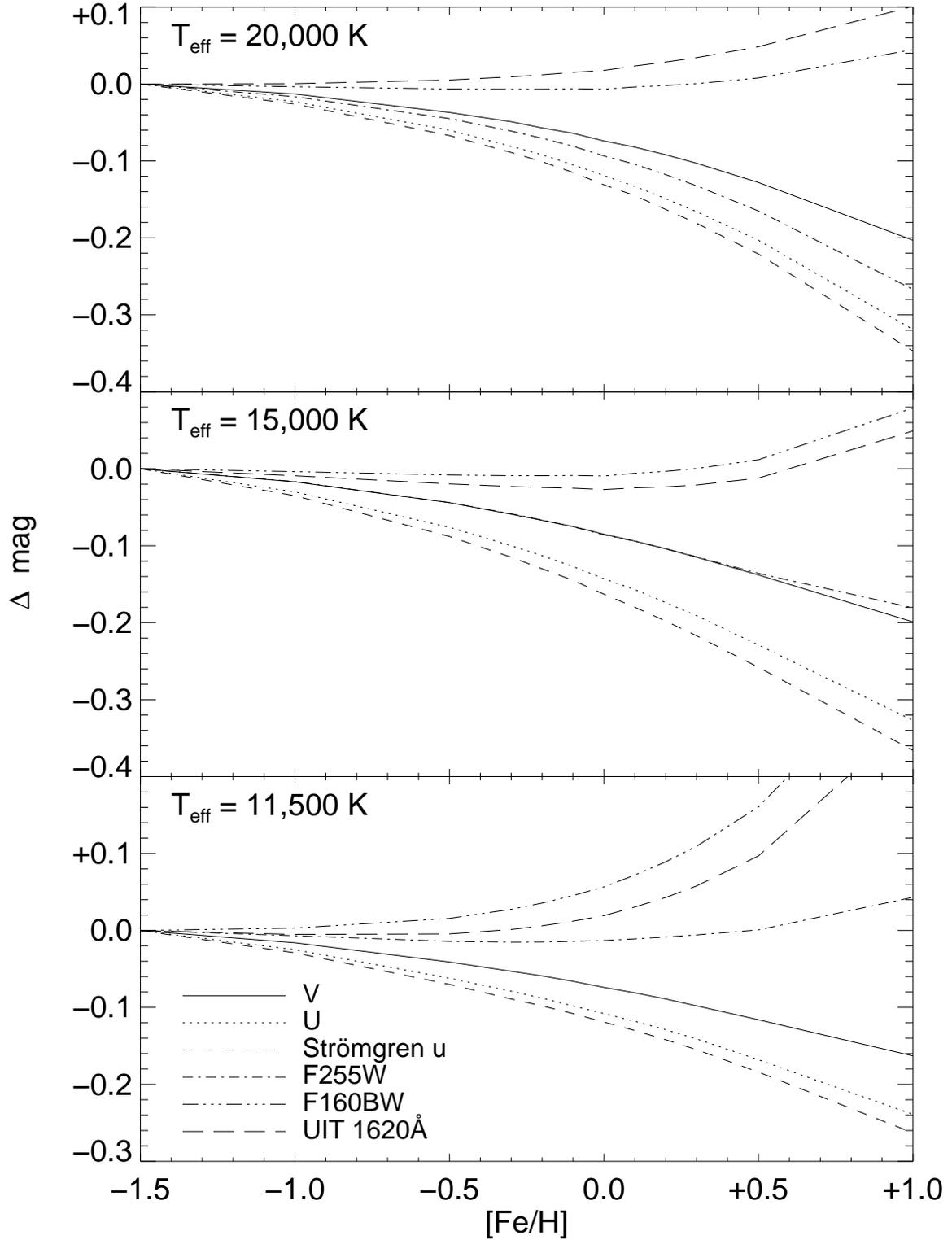}
\caption{\footnotesize The emergent flux in different bandpasses is shown 
          as a
          function of metallicity for a Kurucz model atmosphere with
          $T_{\rm eff} = 16,\!000$~K and $\log\,g = 4.0$. Fluxes are
          normalized so that all bandpasses have a magnitude of 0.0 at
          ${\rm [Fe/H]} = -1.5$. The bandpasses include the Johnson
          $V$ and $U$ filters, the \stromu\ filter, the ultraviolet F255W
          and F160BW filters on WFPC2, and the B5 (1620~\AA) UIT filter.
          }
\end{figure*}

%
\begin{figure*}
 \figurenum{8}
 \epsscale{1.650}
 \plotone{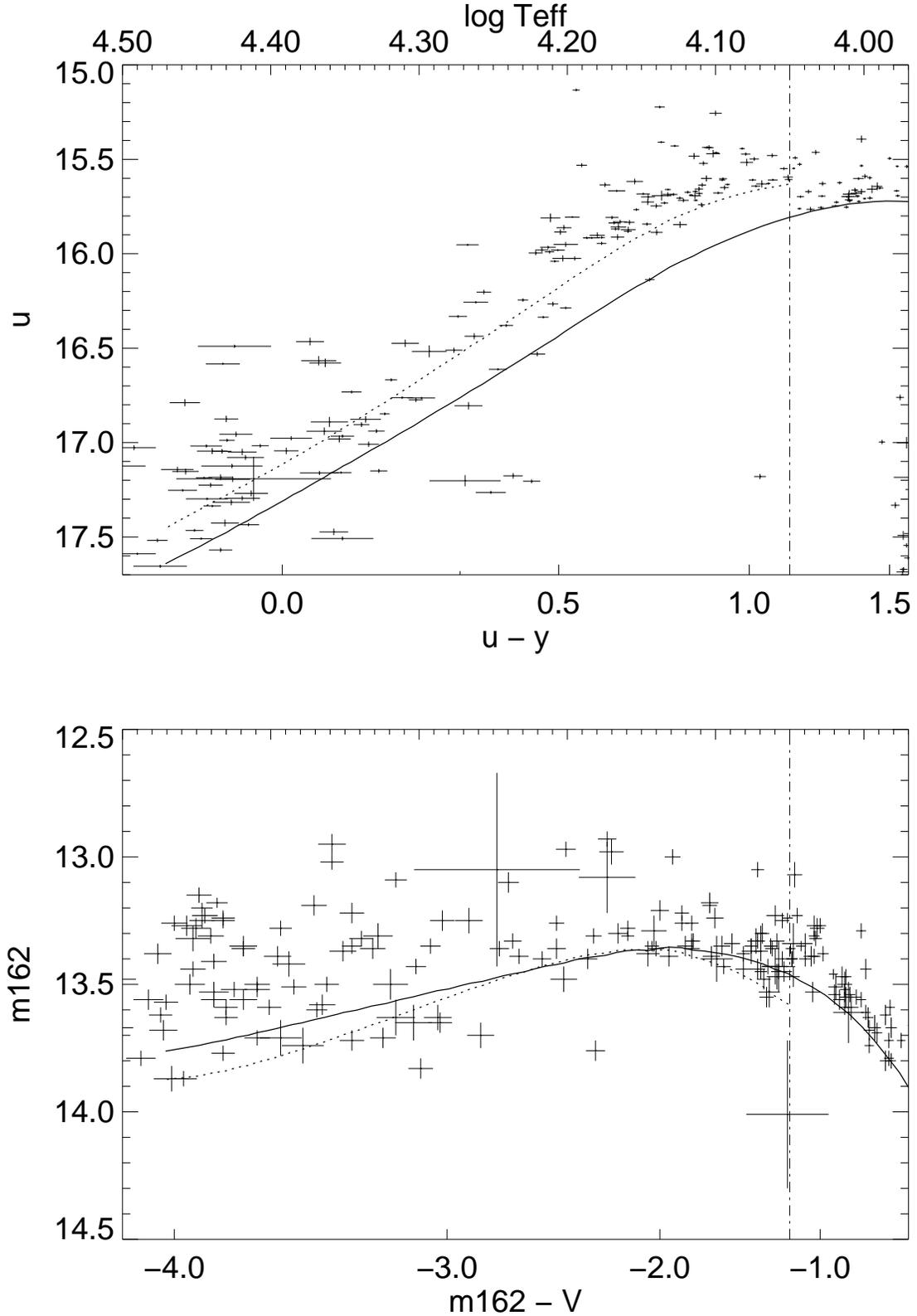}
 \caption{\footnotesize Upper Panel: The $u$,~$u-y$ CMD of NGC~6752 is 
          shown
          along with a canonical ZAHB with [Fe/H]~=~$-1.6$ from 
          Sweigart (see Landsman et al. 1996) transformed to the 
          observational plane using model atmospheres with the cluster 
          metallicity (${\rm [Fe/H]} = -1.5$; solid line) and a 
          suprasolar ([Fe/H]~=~+0.5; dotted line) metallicity. A 
          reddening of $E(\bv) = 0.05$~mag has been assumed. Lower 
          Panel: A similar plot for the  m$_{162}$,~m$_{162} - V$
          CMD of NGC~6752 obtained from the 1620 \AA\  UIT photometry 
          of Landsman et al.\ (1996). The abscissas of both plots have 
          been transformed to a  scale linear in $\log\,T_{\rm eff}$ 
          using Kurucz 
          model atmospheres; this transformation results in the larger
          uncertainties at high $T_{\rm eff}$  in the Str\"omgren $u$ plot. 
          The vertical dot-dash line marks the 
          position ($\log\,T_{\rm eff} = 4.07$) of the Str\"omgren $u$ 
          jump. Note that (1)  there is some evidence for a ``negative
          jump'' to fainter ultraviolet luminosities, and (2) the use 
          of metal-rich atmospheres brightens the ZAHB in the 
          Str\"omgren $u$ CMD, but not in the ultraviolet CMD.
          }      
\end{figure*}

%
\begin{figure*}
 \plotfiddle{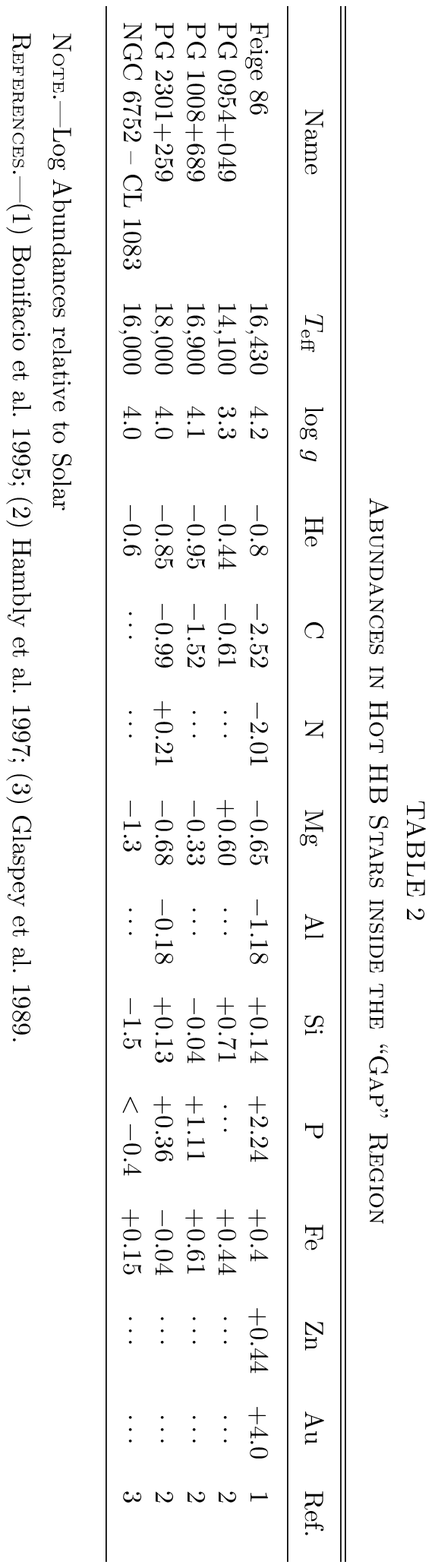}{1.85in}{90}{85}{85}{384}{-116}
\end{figure*}

In conclusion, with the {\em possible} exception of M13, the 
ultraviolet data show {\em no} evidence for a luminosity jump 
corresponding to the jump reported here for \stromu. Interestingly, 
M13 is the cluster for which the strongest evidence for deep mixing 
is currently available (see Table~1).

\section{Levitation of Heavy Elements: a Possible Explanation} 

The \stromu\ bandpass is located just shortward of the Balmer jump 
and thus the emergent flux is dominated by the hydrogen  opacity. 
Atmospheric effects (related, e.g., to an increase in  the metal 
opacity) that decrease the relative importance of the  hydrogen opacity 
should result in a brighter \stromu\ flux.   Figure~7 shows how the flux 
in different bandpasses varies as a  function of metallicity for Kurucz 
models (taken from Lejeune et  al.\ 1997) at three temperatures  
($T_{\rm eff} = 11,\!500$~K, $16,\!000$~K, and  $20,\!000$~K), which span 
the range of the \stromu\ jump.  At all three temperatures, the maximum 
brightening occurs in \stromu\, and at $T_{\rm eff} = 16,\!000$~K  the 
model with ${\rm [Fe/H]} = +0.5$ is about 0.3~mag brighter in \stromu\  
than the model with ${\rm [Fe/H]} = -1.5$.    In contrast, in the 
ultraviolet ($\approx 1600$~\AA) bandpasses, the models with 
${\rm [Fe/H]} = +0.5$ either show little difference, or (at $T_{\rm eff} =
11,\!500$~K) are about 0.1~mag {\rm fainter} than the models  with ${\rm
[Fe/H]} = -1.5$.
As discussed below,
several lines of evidence suggest that radiative levitation can 
enormously enhance the heavy metal abundance in hot HB stars, an 
effect similar to that seen at a similar \teff\ in the Hg-Mn stars 
and other helium-weak, (non-magnetic) chemically peculiar (CP), B-type 
stars (e.g., Dworetsky 1993). {\em We thus suggest that the $u$-jump 
reported here, and its absence in the ultraviolet, is most likely due 
to radiative levitation of heavy elements to supra-solar abundances.}

Figure~8 shows the \stromu\ and ultraviolet CMDs of NGC 6752, with a canonical
ZAHB  from Sweigart (see Landsman et al.\ 1996) transformed to the
observational planes using model atmospheres with the cluster metallicity
(${\rm [Fe/H]} = -1.5$) and with a supra-solar  metallicity
([Fe/H]~=~+0.5).\footnote{In principle, the boundary  conditions used to
compute the interior models should be modified  when using a metal-rich model
atmosphere. However, the implications  of this approximation for the results
described in this paper are  minor.}  In the \stromu\ CMD, for temperatures
hotter than the jump temperature, the  metal-rich model provides a much better
fit than the model with the  cluster metallicity.   The metal-rich model also
provides a somewhat better fit for temperatures hotter than the jump
temperature in the ultraviolet CMD, where there is some evidence for a
``negative jump" to fainter ultraviolet luminosities.   For temperatures cooler
than the jump temperature, radiative levitation presumably does not occur.
The sudden onset of the jump at a 
well-defined temperature, $T_{\rm eff}^{\rm jump} = 11,\!500$~K, 
is possibly a result of the competition between the radiative 
levitation and nuclear (HB) timescales: radiation forces increase 
with $T_{\rm eff}$ so that it is conceivable that there is a ``critical 
temperature" above which radiative acceleration becomes effective in 
a time much shorter than the HB lifetime.\footnote{The phenomenon
could be, to a smaller extent, also connected to the  rotational 
velocities of HB stars, in the sense that HBs containing faster 
rotators might be able to inhibit the onset of radiative levitation 
until a slightly higher temperature is achieved. In fact, this may 
provide an explanation for the (small) difference in 
$T_{\rm eff}^{\rm jump}$ between M3 and M13 (\S3), since it is 
well known that HB stars in M13 rotate significantly faster than   
their M3 counterparts (Peterson 1983; Peterson et al.\ 1995).}

At the high temperatures ($\log\,T_{\rm eff} > 4.3$) and gravities 
of the sdB (EHB) stars, the  metal-rich models in Figure~8 are too 
bright in Str\"{o}mgren $u$, which probably indicates that radiative 
levitation is no longer as effective. Bergeron et al.\ (1988) posited 
the existence of additional transport processes in sdB atmospheres, 
such as a weak stellar wind, to explain why silicon abundances were 
{\em observed} to be strikingly lower than predicted by radiative 
levitation models (also see Fontaine \& Chayer 1997). Studies of the 
pulsation modes in sdB stars suggest that radiative levitation of 
iron occurs in these stars (Charpinet et al.\ 1997), {\em but} not 
necessarily reaching the photosphere.

An important caveat in the interpretation of Figures~7 and 8 is that 
hot HB stars are known to have helium depletions (e.g., Moehler et 
al.\ 1995, 1997) and (as discussed below)  likely do {\em not} show 
significant enhancements of most of the light (${\rm A} \lesssim 34$) 
elements. Both of these effects will somewhat reduce the brightening 
in \stromu\ predicted by the Kurucz models, which use a solar helium 
abundance and solar-scaled metallicities. In addition, the predicted 
ultraviolet fluxes are uncertain if the important carbon and silicon 
opacity sources do not scale with the heavy metals. Better predictions 
of the flux distribution in hot HB stars will most likely require the 
computation of model atmospheres with non-scaled solar abundances, 
for example, by use of the opacity-sampled {\sc atlas12} program 
(Kurucz 1993).   

What is the evidence that significant radiative levitation of 
heavy elements occurs in hot HB stars? First, we note that among 
the main-sequence B- and A-type stars, slow 
($v \sin i < 80\, {\rm km}\,{\rm s}^{-1}$) rotation appears to be a 
necessary condition for the appearance of abundance peculiarities 
(Wolff \& Preston 1978; Abt \& Morrell 1995). Although \vsini\ 
measurements of hot HB stars are not yet available, observations 
of somewhat cooler BHB stars yield upper limits of 
$v \sin i \lesssim 40\,{\rm km}\,{\rm s}^{-1}$, and no indication 
for an increase in \vsini\ with $T_{\rm eff}$ (Peterson et al.\ 
1995; Cohen \& McCarthy 1997). The observed helium depletions provide 
more direct evidence that chemical separation is possible in hot HB
stars. On the theoretical side, the calculations of radiative 
levitation and diffusion processes in hot HB stars by Michaud, 
Vauclair, \& Vauclair (1983) indicate that if the outer envelope 
is stable enough for the gravitational settling of helium to be 
efficient, then overabundances of heavy elements by factors of 
$10^3 - 10^4$ are expected.            

Direct evidence for radiative levitation of heavy elements comes 
from the echelle spectroscopy of two hot HB stars in NGC~6752 by  
Glaspey et al.\ (1989). An overabundance of iron by a factor of  
50 (and a helium depletion) was found in the star CL~1083 with  
$T_{\rm eff} = 16,\!000$~K (within the \teff\ range of the jump).  
On the other hand, no abundance anomalies were found in the star  
CL~1007, which at $T_{\rm eff} = 10,\!000$~K lies coolward of  
$T_{\rm eff}^{\rm jump}$. Similarly, Lambert, McWilliam, \& Smith 
(1992) obtained high-resolution spectra of three cluster HB stars 
[two in M4 (NGC~6121) and one in NGC~6397] located coolward of the 
jump at $T_{\rm eff} \sim 9000$~K, and found no abundance anomalies. 
Unfortunately, there has been no further  echelle spectroscopy of hot 
GC HB stars to confirm the Glaspey et  al.\ result, and to explore the 
prevalence and temperature range of supra-solar iron abundances in hot 
HB stars.\footnote{After this paper was submitted, a preprint became 
available reporting on Keck spectroscopy of BHB stars in M13 (Behr et 
al. 1999) which effectively {\em verifies the Glaspey et al. results 
and the radiative levitation scenario laid out in the present section}. 
Note that the onset of radiative levitation, as derived from the Behr 
et al. work (their Fig.~1), coincides to a remarkable degree of 
accuracy with $T_{\rm eff}^{\rm jump}$ as determined in our \S3 (see 
also Table~1).   An even more recent (but less accurate) spectroscopic 
analysis of BHB stars in NGC~6752 (Moehler et al. 1999) has also confirmed the 
enhanced Fe (but ``normal" Mg) pattern discussed in this section.}
However, some additional guidance can be provided by high-dispersion 
analysis of helium-depleted field HB stars within the temperature range of the
$u$-jump. Table~2 shows the results of abundance 
analyses for the field HB stars Feige 86 (Castelli, Parthasaraty, 
\&  Hack 1997), PG~0954+049, PG~1008+689, PG~2301+259
(Hambly et al.\ 1997) along with the Glaspey  et al.\ result for the 
cluster HB star NGC~6752~--~CL~1083. Not shown in Table~2 are the 
results of Heber (1991), who did not perform a  full abundance 
analysis, but does report chlorine abundances, respectively, 
enhanced by factors of twenty and forty over solar for the BHB stars 
PHL~25 ($T_{\rm eff} = 19,\!000$~K; Ulla \& Thejll 1998) and PHL~1434   
($T_{\rm eff} = 19,\!000$~K; Kilkenny \& Busse 1992). In general, the 
hot HB stars show depletions of helium and the light elements (with 
the exceptions of chlorine and phosphorous), but supra-solar 
abundances of iron and heavier elements. Of course, one does not 
know the original abundances of the field hot HB stars, but 
observations of somewhat cooler field HB stars do suggest that they 
arise from a metal-poor population. For example, Gray et al.\ (1996) 
find that field HB stars between 7000~K  and 9000~K are metal-poor 
(less than ${\rm [Fe/H]} \sim -1$) and lie  close to the ZAHB in the 
$\log\,g$,~$\log\,T_{\rm eff}$ diagram.

As noted above, the abundances of most of the light metals in hot HB stars 
do not seem to show the same enhancement as the heavy metals. This effect 
can be understood in terms of two circumstances which preferentially favor 
saturation of the radiative forces in the light elements: the light metals 
generally have larger initial abundances and a less rich absorption spectrum 
(a few strong lines rather than many weak lines) than the heavy metals. The 
absence of overabundances in the light elements will make it difficult to 
detect the presence of radiative levitation in low-dispersion optical and 
ultraviolet spectra.  The strongest lines in low-resolution optical spectra 
of hot HB stars are due to ions of the light elements such as \ion{C}{2}, 
\ion{Mg}{2}, and \ion{N}{2}. Similarly, in the ultraviolet the strongest 
lines are due to ions of the light elements, although in this case, one 
expects the continuum to be depressed by the presence of numerous weak 
iron-peak lines.   Such a depression of the far-UV continuum might have been
seen by Vink et al.\ (1999), who analyzed a far-UV spectrum of M79 obtained with
the Hopkins Ultraviolet Telescope (HUT).    They suggest that the  
the agreement between their
synthetic and the observed  spectrum could be improved if the surface abundances
of the hot HB stars in M79 were enhanced by radiative levitation.     

IUE spectra of GC HB stars (Cacciari et al.\ 1995)  
do not show any especially strong absorption features, and, in particular, 
do not show the strong \ion{Si}{2} photoionization resonances, which 
dramatically distort the far-ultraviolet continuum in the ApSi stars 
(Lanz et al.\ 1996). Thus, silicon is almost certainly not enhanced to 
suprasolar abundances in GC BHB stars. 

As we have seen previously (\S4), the $u$-jump is strongly 
correlated with the $\log\,g$-jump. What is the effect of radiative 
levitation on the derived gravities of hot HB stars? 

The gravities are derived by finding the gravity of a model (of a
given temperature and metallicity) which best fits the Balmer line 
profiles (e.g., Saffer et al.\ 1994; Moehler et al.\ 1995). The 
temperature must be either determined independently (e.g., from the 
ultraviolet continuum), or the gravity and temperature can be 
determined together from simultaneous fitting of multiple Balmer 
lines. Thus, to determine how radiative levitation of heavy elements 
can alter the derived gravity, one must also consider how the 
temperature is derived. This exercise was performed by Moehler et 
al., who compared derived gravities for hot HB stars in M15 using 
models with both solar and $0.01 \times\,{\rm solar}$ metallicity 
(close to the cluster metallicity). They found that gravities could 
be underestimated by at most 0.1~dex if the HB stars had solar 
metallicities and metal-poor models were used to analyze them.  
They concluded that radiative levitation was insufficient to
explain the size ($\sim 0.2$~dex) of their observed low gravity 
anomaly (the $\log\,g$-jump). In fact, their exercise is 
consistent with our \stromu\ study in that it requires that 
heavy element abundances must be significantly {\em above} solar, 
in order for radiative levitation to be the origin of the anomaly. 
This statement is supported by the study of Leone \& Manfr\`e 
(1997) who, in their analysis of helium-weak stars, found that the 
derived \logg\ value could be underestimated by up to 0.25~dex if 
a solar metallicity model were used to determine the gravity of 
helium-weak stars with a heavy metal abundance ten times solar. 

In addition to low gravities, the spectroscopic studies of de Boer 
et al.\ (1995) and Moehler et al.\ (1995, 1997) led to HB masses 
(derived from values of the stellar \teff, \logg, $V$ magnitude, and 
the cluster distance) significantly below canonical values. Heber, 
Moehler, \& Reid (1997) found that this discrepancy could be partially 
alleviated by use of the larger cluster distances indicated by some 
{\sc Hipparcos} studies (e.g., Reid 1997; Gratton et al.\ 1997), 
although the derived masses were still lower than canonical values 
for NGC~6397 and NGC~288. The use of the long distance scale also 
led to absolute magnitudes brighter than canonical models, leading 
Heber et al.\ to favor non-canonical evolutionary models. However, 
if our hypothesis of radiative levitation is correct, then the 
derived masses must be considered uncertain at best, at least for HB 
stars in the ``critical" temperature range, 
$11,\!500~{\rm K} \lesssim T_{\rm eff} \lesssim 20,\!000$~K. 
 
Recently, Caloi (1999) has also proposed that radiative levitation occurs 
in hot HB stars, mainly based on the suggested existence of a gap in the 
HB number counts in several GC CMDs near $\bv \sim 0.0$~mag. In principle, 
such gaps could be related to the ``jumps'' discussed in this paper; for 
example, if a luminosity jump were much more prominent in $B$ than in $V$, 
then a gap would appear at the location of the onset of the jump in a 
$V$,~$\bv$ diagram---but {\em only if} $T_{\rm eff}^{\rm jump}$ could be 
associated to a $\bv$ color along the ``horizontal" part of the HB. However, 
the temperature corresponding to $\bv = 0.0$~mag is $\approx 8500$~K, 
much cooler than the $11,\!500$~K found here for the onset of the \stromu\ 
jump. In addition, it appears that a gap at $\bv = 0.0$~mag is not a 
{\em ubiquitous} phenomenon (see, e.g., the Appendix in Catelan et al.\ 
1998), contrary to what might be expected in Caloi's scenario. Finally, 
we also note that our $u$,~$y$ data for M68 (NGC~4590) from ESO do not 
show clear evidence for a $u$-jump because its {\em hottest} BHB stars 
are close to the temperature limit for the onset of the jump. In summary, 
it is unlikely that the gaps discussed by Caloi are related to the jump 
discussed in this paper, although the connection between HB gaps and 
atmosphere effects merits further investigation.

To summarize, radiative levitation of heavy elements can plausibly 
explain the temperature range and magnitude of both the 
$u$- and $\log\,g$-jump as well as the low-mass problem, but 
further high-resolution optical and ultraviolet spectra are needed
to demonstrate that the abundances of iron and other heavy elements are 
significantly above solar. As part of this effort, we have a 
current HST Cycle~8 program (GO-8256) to obtain STIS ultraviolet 
spectra of nine HB stars in NGC~6752 which span the temperature
range of the jump.   Also needed are model atmospheres with
non-scaled solar abundances (computed, e.g., with the {\sc atlas12} 
code) to  determine quantitatively whether the observed \stromu\ 
and gravity anomalies can be entirely explained by overabundances 
of heavy elements or whether additional effects such as those 
discussed in \S5 (i.e., helium mixing) are required. In this 
regard, we issue a cautionary remark on attempts to calibrate 
the free parameters of the helium-mixing theory (which is not a 
``first-principles" theory) using the $u$- and $\log\,g$-jump 
properties: before this task can be successfully accomplished, the 
effects of radiative levitation upon the adopted model atmospheres 
must be taken into account. 

Finally, we note that there have been no theoretical studies of 
diffusion processes in hot HB stars since the work of Michaud et 
al.\ (1983), and that much more sophisticated calculations of 
radiative accelerations are now possible (e.g., Richer et al.\ 1998).

\section{Summary and Concluding Remarks} 

In the present paper, we have carried out an extensive analysis 
of the Grundahl et al.\ (1998) ``jump" in \stromu\, first detected 
in M13. With this purpose, we presented new $u$,~$y$ photometry 
of fourteen GCs based on four filter/detector combinations.

The main results of our analysis of this large set of $u$,~$u-y$
CMDs can be summarized as follows:

 1.~The \stromu\ jump is a ubiquitous feature, present in every 
       metal-poor GC with sufficiently hot BHB stars. Such a jump  
       is morphologically best described as a systematic deviation, 
       in $u$ magnitudes and/or $u-y$ colors, with respect to the 
       expectations of canonical ZAHB models, in the sense that 
       the observations appear brighter and/or hotter than the 
       theoretical predictions; 

 2.~The parameter that best defines the onset of the jump is 
       its temperature, which we find to be remarkably constant 
       from one cluster to the next: 
       $T_{\rm eff}^{\rm jump} = 11,\!500 \pm 500$~K; the error 
       estimate is essentially due to measurement and/or calibration
       uncertainties. We do not find any significant evidence for 
       a dependence of $T_{\rm eff}^{\rm jump}$ on metallicity. 
       The high-temperature end of the jump appears to be situated 
       at $\approx 20,\!000$~K; 

 3.~The occurrence of the jump is not related to the GC 
       metallicity, central concentration, central density, 
       extent of mixing on the RGB, HB morphology (provided the 
       ``critical" temperature $T_{\rm eff}^{\rm jump}$ is 
       reached by the cluster's BHB), or Galactocentric distance;  

 4.~The height (or ``size") of the $u$-jump is remarkably   
       constant amongst our entire sample of GCs; 

 5.~The $u$-jump is intimately connected, on a star-by-star 
       basis, to the low gravities ($\log\,g$-jump) which have 
       been measured for GC BHB stars.

Recently, a non-canonical evolutionary scenario (helium mixing:
Sweigart 1997a, 1997b) has been proposed as a possible explanation 
for the low BHB gravities ($\log\,g$-jump)---which, as we have just
remarked, seems strongly connected to the $u$-jump. From our 
discussion, we were able to pose the following constraints on this 
scenario:

 1.~{\em ``Global" fine-tuning problem}: Given the strong 
       dependence of ZAHB properties upon variations in GC 
       evolutionary parameters, one would naturally expect 
       some intrinsic relationship between 
       $T_{\rm eff}^{\rm jump}$ and [Fe/H]. However, 
       any intrinsic relationship between these two 
       quantities, if present at all, seems to be quite 
       mild---posing, in fact, a major challenge for {\em any} 
       stellar evolution-related scenario for the occurrence 
       of the jump, and pointing instead to a stellar 
       atmospheres-based solution; 

 2.~{\em ``Local" fine-tuning problem}: (Extreme) fine 
       tuning is also required in the helium mixing scenario 
       {\em at any given metallicity} and {\em for any given GC} 
       in order for $u$- and $\log\,g$-jumps such as the ones 
       observed to be reproduced by the non-canonical models;  

 3.~Helium mixing theory predicts that the jump size and 
       location should depend quite strongly on the extent of 
       deep mixing on the RGB. However, GCs in which the RGB 
       stars have undergone extreme deep mixing---such as 
       M13---present jump characteristics virtually identical 
       to those of GCs whose giants seem to have undergone 
       little mixing---such as NGC~288; 

 4.~If (as commonly assumed) deep mixing on the RGB is related 
       to stellar rotational velocity, current measurements of HB 
       rotational velocities (Peterson et al.\ 1995) would lead 
       one to expect (perhaps large) intrinsic scatter in 
       $T_{\rm eff}^{\rm jump}$ {\em and} jump size at any given 
       metallicity---contrary to what our observations appear to 
       suggest;  

 5.~The jump phenomenon (at least in $\log\,g$) is present 
       not only among GC BHB stars, but also in the field 
       (e.g., Mitchell et al.\ 1998). Since it is well known 
       that RGB stars in the field do not show deep mixing 
       patterns nearly as large as GC giants (e.g., Hanson et 
       al.\ 1998; Carretta et al.\ 1999b; Kraft 1998, 1999), 
       their progeny must clearly not have undergone helium mixing.  
       This provides strong indication that deep mixing cannot 
       be responsible for the jump phenomenon. In addition, it 
       may also imply that (most) EHB (sdB) stars in the {\em halo} 
       field (Mitchell 1998), and possibly also in GCs, cannot have 
       their origin ascribed to helium mixing on the RGB; 

 6.~With the {\em possible} exception of M13, the jump 
       phenomenon is {\em not} seen in ultraviolet CMDs, and thus 
       does {\em not} appear to be caused by a jump in the 
       bolometric luminosity---contrary to what would be expected 
       in the helium-mixing scenario.

These observations suggest that a stellar atmosphere effect, 
rather than helium mixing, is the primary cause of the $u$- and 
$\log\,g$-jump phenomenon. {\em We propose here that radiative 
levitation of metals might be able to explain all aspects of the 
jump problem.} This suggestion, which requires further development 
on the basis of new observations and diffusion/model atmosphere 
computations, is in essence based on the following main lines of 
evidence:

 1.~The temperature range of the jump is similar to that found for 
       the chemically peculiar (Hg-Mn and helium-weak) B-type stars, 
       which show helium depletions and large overabundances of heavy 
       elements. Observations of (somewhat cooler) BHB stars show them 
       to be slow  ($v \sin i \lesssim 40\, {\rm km}\,{\rm s}^{-1}$) 
       rotators (Peterson et al.\ 1995; Cohen \& McCarthy 1997), 
       and slow rotation ($v \sin i < 80\, {\rm km}\,{\rm s}^{-1}$) 
       seems to be a necessary condition for the appearance of
       overabundances in the B-type stars (Wolff \& Preston 1978). The 
       helium depletions observed in the hot HB stars (Moehler et al.\ 
       1995) show that chemical separation is feasible in these stars. 
       Theoretical considerations (Michaud et al.\ 1983) suggest that if 
       an HB atmosphere is stable enough to show helium depletion, then 
       overabundances of heavy metals by factors of $10^3 - 10^4$ might 
       be expected;     

 2.~An abundance analysis derived from echelle spectra of the star 
       CL~1083 (\teff~$= 16,\!000$~K) in NGC~6752 yielded an 
       overabundance of iron by a factor of 50 (Glaspey et al.\ 1989) 
       and observations of field HB stars within the temperature 
       range of the jump consistently show an overabundance of iron-peak
       and heavier metals (Table~2);  

 3.~Simple experiments with Kurucz model atmospheres suggest that an
       increase of the metallicity to suprasolar abundances can lead to 
       0.3~mag brightening of the \stromu\ flux, with little change or a 
       decrease in the ultraviolet flux. The work of Leone \& Manfr\`e 
       (1997) suggests that an underestimate of the gravity by as much as 
       0.25~dex might result if super metal-rich spectra were analyzed 
       using models with the cluster metallicity. This implies that  
       efforts to employ $u$,~$y$ or $\log\,g$,~$\log\,T_{\rm eff}$ 
       diagrams to constrain non-canonical evolutionary models cannot be 
       reliably carried out until the effects of radiative levitation 
       upon the adopted model atmospheres have been properly taken into 
       account.

Our scenario, as laid out above, leads to several predictions, which we 
encourage observers to test. Among these predictions, we may highlight 
the following:

 1.~Every metal-poor GC with a sufficiently long blue tail will show 
       the jump phenomenon;  

 2.~Even $\omega$~Cen will show a well-defined jump in $u$ 
       and in $\log\,g$, in spite of its large intrinsic spread 
       in metallicity (by $\sim 1$~dex: e.g., Norris, Freeman, \& 
       Mighell 1996; Suntzeff \& Kraft 1996). Moreover, the 
       low-temperature cutoff of the jump is predicted to be located
       at the same place in both $\omega$~Cen and NGC~288 
       (i.e., $T_{\rm eff}^{\rm jump} = 11,\!500 \pm 500$~K), in 
       spite of the dramatic differences in mixing history between 
       the two globulars ($\omega$~Cen: Norris \& Da Costa 1995a, 
       1995b; Zucker, Wallerstein, \& Brown 1996; NGC~288: 
       see Table~1);       

 3.~Any bona-fide BHB star---whether in GCs or in the 
       field---lying in the critical temperature range 
       $11,\!500~{\rm K} \lesssim T_{\rm eff} \lesssim 20,\!000$~K,
       will lie above the canonical ZAHB loci in the $u$,~$u-y$ 
       and $\log\,g$,~$\log\,T_{\rm eff}$ planes;   

 4.~The radiative levitation hypothesis will be easily falsifiable, 
       once additional echelle spectra of cluster hot HB stars have 
       been obtained. (This project is feasible for the nearest 
       GCs using the coming generation of 8-m and larger telescopes.) 
       Should the derived iron abundances not be consistently above 
       solar, then an alternative explanation will be required for the 
       $u$- and $\log\,g$-jump phenomenon. On the other hand, if the 
       suprasolar iron abundances are confirmed, then metal-rich model 
       atmospheres (with non-solar-scaled abundances) must be constructed 
       to derive the fundamental stellar parameters.

Our GC sample is comprised of inner-halo clusters only. (For the 
HB morphology-independent definition of ``outer halo," the reader 
is referred to \S7 in Borissova et al.\ 1997 and references 
therein.) It would prove of interest to investigate whether the 
jump is present in outer-halo GCs---NGC~6229 being an ideal 
candidate for further examination (Borissova et al.\ 1999)---and 
in bulge GCs with blue HBs (Ortolani, Barbuy, \& Bica 1997 and 
references therein). 

As pointed out in \S5, there seems to be a significant correlation 
between deep-mixing signatures on the RGB of GCs and HB morphology. 
{\em If} helium mixing should turn out not to be the cause, whence 
this correlation? 
One possibility is that non-canonical mixing on the RGB is 
related to mass loss (through stellar rotation?). This idea
has tentavile been raised by Catelan \& de Freitas Pacheco
(1995) and Kraft et al. (1995).
In this regard, we would 
like to mention that whereas ``virtually all giants in M13 mix as 
they approach the red giant tip" (Kraft 1998; see Fig.~10 in Kraft 
1994), redward of the jump---where $\approx 50\%$ of all M13 HB 
stars are found---oxygen abundances appear to be ``normal" (see 
Fig.~7 in Peterson et al.\ 1995). This is obviously a very 
surprising result: where are the RGB progenitors of such BHB stars 
in M13? Could the discrepancy be related, at least in part, to 
Langer's (1991) mass loss scenario, whereby (some) RGB stars might 
appear more oxygen-poor than they actually are due to forbidden 
\ion{O}{1} emission from an extensive, cool, slowly expanding outer 
envelope---possibly implying somewhat enhanced mass loss rates? We 
note that Langer's hypothesis has thus far neither been conclusively 
ruled out nor corroborated (see, e.g., Minniti et al.\ 1996 for a 
recent discussion).  

It should also be extremely interesting to investigate the position 
of BHB stars in the $u$,~$u-y$ and $\log\,g$,~$\log\,T_{\rm eff}$ 
planes in the mildly metal-rich (${\rm [Fe/H]} \approx -0.5$~dex) 
GCs NGC~6388 and NGC~6441. As discussed by Sweigart \& Catelan 
(1998) and Layden et al.\ (1999), these two globulars represent 
extreme examples of the second-parameter phenomenon. All of the 
theoretical scenarios laid out by Sweigart \& Catelan predict 
anomalously bright HB stars, thus implying {\em intrinsically} low 
gravities and $u$ magnitudes, even in regions outside the $u$- and 
$\log\,g$-jump. Preliminary results from Moehler, Sweigart, \& 
Catelan (1999) have indicated surprisingly high gravities for BHB 
stars in these clusters, although more data appear to be needed to 
confirm such high gravities. 

\acknowledgments
The authors would like to thank R. A. Bell, B. F. W. Croke, B. Dorman, 
F. R. Ferraro, R. P. Kraft, T. Lanz, S. Moehler, B. Paltrinieri, 
C. R. Proffitt, R. T. Rood, R. A. Saffer, M. D. Shetrone, A. V. 
Sweigart, and D. A. VandenBerg for helpful information and/or 
discussions. We are also grateful to the anonymous referee whose 
comments and suggestions led to a significant improvement in the 
presentation of our results. F.G. gratefully acknowledges financial 
support from the Danish Natural Sciences Research Council and 
Don A. VandenBerg. He also acknowledges the hospitality and financial 
support offered by the National Research Council of Canada for making 
his stay at the Dominion Astrophysical Observatory possible. This 
research was supported by the Danish Natural Science Research Council 
through its Centre for Ground-Based Observational Astronomy (IJAF). 
Support for M.C. was provided by NASA through Hubble Fellowship grant 
HF--01105.01--98A awarded by the Space Telescope Science Institute, 
which is operated by the Association of Universities for Research in 
Astronomy, Inc., for NASA under contract NAS~5--26555. This research 
has made use of archived data from the Canadian Astronomy Data Centre 
(CADC), which is operated by the Herzberg Institute of Astrophysics, 
National Research Council of Canada.

\end{document}